\documentclass[12pt]{amsart}

\newenvironment{prf}{\noindent{\it Proof\/}:}{$\;\square$
\par\medskip}

\newtheorem%
{thm}{Theorem}[section]
\newtheorem%
{proposition}[thm]{Proposition}
\newtheorem%
{lemma}[thm]{Lemma}
\newtheorem%
{lemmadef}[thm]{Lemma-Definition}
\newtheorem%
{corollary}[thm]{Corollary}
\newtheorem%
{conjecture}[thm]{Conjecture}

\newcommand{\dontprint}[1]{\relax}

\def\gh{\mathrm{gh}}
\title
[Lagrange structure and quantization] {Lagrange structure and
quantization}
\author{P.O. Kazinski, S.L. Lyakhovich and  A.A. Sharapov}
\address{Department of Quantum Field Theory, Tomsk State University, Tomsk 634050, Russia}
\email{kpo@phys.tsu.ru, sll@phys.tsu.ru, sharapov@phys.tsu.ru}

\thanks{We are thankful to
Jim Stasheff for many useful comments on the manuscript. This work
was partially supported by the RFBR grant 03-02-17657, Russian
Ministry of Education grant 130337, and the grant for Support of
Russian Scientific Schools 1743.2003.2. AAS and POK appreciate
financial support from the Dynasty Foundation and International
Center for Fundamental Physics in Moscow.}

\begin{document}

\begin{abstract}
A path-integral quantization method is proposed for dynamical
systems whose classical equations of motion  do \textit{not}
necessarily follow from the action principle. The key new notion
behind this quantization scheme is the Lagrange structure which is
more general than the Lagrangian formalism in the same sense as
Poisson geometry is more general than the symplectic one. The
Lagrange structure is shown to admit a natural BRST description
which is used to construct an AKSZ-type topological sigma-model.
The dynamics of this sigma-model in $d+1$ dimensions, being
localized on the boundary, are proved to be equivalent to the
original theory in $d$ dimensions. As the topological sigma-model
has a well defined action, it is path-integral quantized in the
usual way that results in quantization of the original (not
necessarily Lagrangian) theory. When the original equations of
motion come from the action principle, the standard BV
path-integral is explicitly deduced from the proposed quantization
scheme. The general quantization scheme is exemplified by several
models including the ones whose classical dynamics are not
variational.
\end{abstract}

\dontprint{
We prove }

\maketitle


\section{Introduction}

We suggest a method for path integral quantization of classical
theories whose equations of motion are not necessarily
variational. The key idea behind the method is that any classical
dynamics can be uniformly converted into an equivalent topological
field theory based on the action principle. Below in the
introduction we informally comment on the main ingredients of the
construction to give a preliminary impression of the quantization
method we propose.

In the first place, we introduce a notion of \emph{Lagrange
structure} that generalizes the standard Lagrangian formalism more
or less in the same sense as the Poisson geometry generalizes the
symplectic one. The  Lagrange structure does not require the
equations of motion to be Lagrangian in usual sense, i.e. no
integrating multiplier is assumed to exist bringing the equations
to the variational form. The main ingredient of the Lagrange
structure is the \emph{Lagrange anchor} (denoted by $V$) that
would be the inverse to an integrating multiplier $\Lambda$ if the
latter existed. The Lagrange anchor is required to satisfy a chain
of compatibility conditions involving the equations of motion and
the gauge generators for the gauge systems. If the anchor is
invertible, these conditions will be equivalent to that providing
the inverse matrix $\Lambda = V^{-1}$ to be an integrating
multiplier for the equations of motion. Given the Lagrange
structure, one can perform the path integral quantization of the
classical theory even when the anchor is not invertible, and
therefore, the dynamics do not admit any action functional.

Another explanation for the origin of the Lagrange structure is
provided by the BV formalism \cite{BV}, \cite{HT}. The standard BV
method describes the (gauge) system in terms of field-antifield
supermanifold endowed with a master action and the canonical
antibracket. Due to the Jacobi identity, the corresponding
antibracket is automatically compatible with the BRST differential
(associated with the master action) in the sense of the Leibnitz
rule. For this standard case, the Lagrange anchor appears in the
theory as an odd bivector defining canonical antibracket between
fields and anti-fields. Similarly, as we show, the generic
Lagrange structure equips an appropriately superextended original
space with a BRST differential and a (non-canonical, weak)
antibracket. The BRST differential carries all the information
about the original classical theory, whereas the antibracket
contains the ingredients needed for quantization. To make the
formalism working (i.e. sufficient for the path integral
quantization), the antibracket is required to satisfy the graded
Jacobi identity in a ``weak" sense, i.e. up to homotopy w.r.t. the
BRST differential. In this algebraic setting, all the
compatibility conditions between classical equations of motion and
Lagrange anchor are encoded in the graded Leibnitz rule for the
BRST differential and the weak antibracket. As the antibracket can
be degenerate, the BRST differential is not necessarily
anti-hamiltonian vector field, in contrast to the standard
Lagrangian theory.

Regarding the relaxed Jacobi identity for the weak antibracket,
the Lagrange structure can be considered as Lagrangian counterpart
of the \textit{weak Poisson structure} studied in \cite{LS1}.
Various particular types of the weak Poisson brackets were studied
earlier in \cite{BT}, \cite{DGL}, \cite{BM} and recently in
\cite{BGL1}. The deformation quantization of weak Poisson
manifolds was given in Ref. \cite{LS1} making use of
superextension to the Kontsevich formality theorem
\cite{Kontsevich}, which had not been proved at that moment. An
exhaustive proof has been given to the superformality theorem in
the recent paper \cite{CaFe1} in connection with the deformation
quantization of $P_\infty$-structure (=weak Poisson structure).

  From the pure algebraic viewpoint, the defining relations for
the Lagrange structure can be regarded as structure relations for
$L_\infty$-algebra \cite{HiS}, \cite{SL}. This $L_\infty$-algebra
is a particular example of the homotopy analog of Schouten (= odd
Poisson, Gerstenhaber) algebras studied in Ref. \cite{Vo} in the
context of higher derived brackets. The higher (derived) brackets
were also studied in various related contexts in Refs. \cite{HDB}.

As a next step towards quantization, we work out a BRST
description for the Lagrange structure of a generic (i.e. not
necessarily Lagrangian) classical theory. Making use of this BRST
formalism, we construct a topological sigma-model along the lines
of the AKSZ approach \cite{AKSZ}  and its further
development\footnote{There is a great number of works devoted to
the topological sigma-models. We mention only those references
which are the most relevant for the construction we use.}
\cite{GD}, \cite{CaFe}, \cite{CaFe2}, \cite{BM}. The master
equation for the sigma-model action reproduces the defining
relations for the Lagrange structure of the original theory much
similar as the master equation for the Poisson sigma-model
reproduces the Jacobi identity for the Poisson bivector \cite{I},
\cite{SS}. By construction, the dynamics of the topological
sigma-model is variational and we prove it is equivalent to the
dynamics of the original classical theory. Quantizing this
topological field theory one gets an arbitrary (not necessarily
Lagrangian) original theory quantized. If the original theory had
an action, the transition amplitude for the sigma-model can be
explicitly integrated out in the bulk, resulting in the standard
BV answer for the path integral of the original theory, with the
BV master action built in a usual way from the original action.

Notice that the trivial Lagrange anchor $V=0$ is always admissible
for any equations of motion. The trivial anchor, however, results
in trivial quantization, with no quantum fluctuations appeared:
The path integral for the topological sigma-model is reduced to
integration only over the classical trajectories of the original
theory. In this sense, the transition amplitudes in
(non-Lagrangian) theories with trivial Lagrange anchor are similar
to the classical transition amplitudes studied by Gozzi \textit{et
al.} \cite{Gozzietall} for theories based on the action principle.
In the general case, the quantum fluctuations will be trivial only
for those degrees of freedom which belong to the kernel of the
Lagrange anchor. The quantization technique we develop is
explicitly covariant and does not require any special coordinate
system adjusted for separating the anchor kernel from the other
degrees of freedom\footnote{Moreover, such an adjusted coordinate
system does not exist if the rank of the anchor varies over the
configuration space. However, this separation, whenever it is
possible, allows for a simple interpretation. For example, if the
degrees of freedom belonging to the kernel of the Lagrange anchor
can be explicitly excluded from the classical equations of motion
it is natural to consider them as auxiliary variables. The
equations will become variational for the remaining degrees of
freedom. The auxiliary degrees of freedom do not fluctuate in this
case, while the other ones are quantized in the usual way with an
action defined on the reduced space.}. In this respect, the
proposed path integral quantization is analogous to the
deformation quantization of Poisson manifolds: the degrees of
freedom from the kernel of the (regular) Poisson bivector
correspond to the center of $\ast$-algebra upon quantization,
remaining ``nonquantized'' in this sense.

Let us comment on the paper composition. In the next Section we
elaborate on the origin of the Lagrange anchor and the related
structures from the viewpoint of classical dynamics. In Sect. 3 we
introduce the notion of the regular Lagrange structure and discuss
its connection with strongly homotopy Schouten algebras
($S_{\infty}$-algebras for short). In Sect. 4, the BRST embedding
is worked out for the Lagrange structure and the existence theorem
is proved for the corresponding master equation. As a byproduct,
we obtain one more geometric interpretation of the Lagrange
structure as an infinitesimal deformation of certain Lagrangian
submanifold in the cotangent bundle over the space of
trajectories.  In Sect. 5, proceeding along the lines of the AKSZ
procedure, we construct a topological sigma-model related to the
BRST complex which have been built for the Lagrange structure in
previous section. Being effectively localized on the boundary, the
dynamics of this topological sigma-model in $d+1$ dimensions are
proved to be equivalent to the original $d$-dimensional theory.
Quantizing the topological sigma model we get the path integral
quantization of the original (non-Lagrangian) dynamics. Finally,
in Sect. 6, we consider several examples illustrating the
quantization formalism proposed. As the \textit{first example}, we
show that the quantization of the usual Lagrangian gauge theory,
being performed by our method, is equivalent to the standard
BV-quantization. Next, as the \textit{second example}, we detail
our quantization method for the general classical theory given by
a set of independent (not necessarily Lagrangian) equations of
motion without gauge symmetry. As the \textit{third example} we
consider the systems with the equations of motion being the
first-order ODEs $\dot{x}{}^i = h^i(x)$. For these systems, we
identify purely algebraic (i.e. containing no time derivatives)
Lagrange anchors with the Poisson bivectors compatible to the
vector $h$. When the Poisson bivector is nondegenerate, the
equations of motion are Hamiltonian and can be derived from the
first-order action functional, while in the degenerate case, these
equations are not necessarily either Hamiltonian, or variational.
However, our method provides a natural embedding of this theory
into the equivalent topological sigma-model, so the theory can be
quantized through this embedding. The \textit{fourth example}
deals with the Maxwell electrodynamics formulated in terms of the
strength tensor of electromagnetic field. The Maxwell equations
for the strength tensor are known to be non-Lagrangian and
linearly dependent. We find an explicitly covariant Lagrange
anchor for these equations and quantize them following our general
prescription. As a result, we obtain a path integral giving
precisely the same transition amplitude as usual Faddeev-Popov
path integral defined in terms of the electromagnetic potentials.
This exemplifies the way in which our quantization method can be
made applicable to non-Lagrangian field theories formulated from
outset in terms of strength tensors, like the high-spin gauge
fields \cite{Va}.

\section{Lagrange structure: a preliminary exposition}\label{prelim}

In this section, we give a down-to-earth explanation to the origin
of the Lagrange anchor and related structures. It is the
structures which are behind the path integral quantization of
dynamical systems whose equations of motion do not necessarily
follow from the action principle. More rigorous consideration of
the subject is given in Sect 3.

In the standard Lagrangian formalism, the mechanical system is
specified by an action functional $S: M\rightarrow \mathbb{R}$
defined on the space of all trajectories  (histories) $M$ over the
configuration space of  the system. The true physical trajectories
are postulated to deliver local minima to $S$ that leads to the
equations of motion of the form
\begin{equation}\label{1}
T_i(x)\equiv\partial_iS(x)=0\,,
\end{equation}
where $x^i$ are local coordinates on $M$. When all the critical
points of $S(x)$ are non-degenerate (and hence isolated) one said
about  a  non-degenerate Lagrangian theory; otherwise there can
exist continuous families of trajectories satisfying the same
boundary conditions, in which case one says about a gauge
invariant (or degenerate) Lagrangian theory.

The equations of motion (\ref{1}) can be understood as defined by
an exact 1-form $T = dS$ on an infinite-dimensional manifold $M$.
An immediate consequence of this interpretation is the Helmholtz
criterion
\begin{equation}\label{dT}
    dT=0\,,
\end{equation}
that is a necessary condition for the equations $T=0$ to come from
the action principle. Due to the Poincar\'e lemma,  the closedness
of $T$ ensures the existence of a local action functional $S_U$
defined on any contractible open set $U\subset M$ such that
$T|_U=dS_U$. If $x_0\in U$ is a solution to $T=0$, one can use
$S_U$ to perform the quasi-classical (= perturbative in $\hbar$)
quantization of the system ``near the classical trajectory
$x_0$''.

So, the standard Lagrangian formalism requires the equations of
motion to satisfy two conditions: (i) they must be components of a
one-form on the cotangent bundle to the space of trajectories, and
(ii) this one-form has to be closed, i.e satisfying the Helmholtz
condition (\ref{dT}). Of course, there are many physically
interesting systems whose equations of motion do not satisfy even
the first condition, not to mention the second one. The first
natural step towards generalizing the Lagrangian formalism is to
replace the cotangent bundle $T^\ast M$, where the equations
(\ref{1}) take the values, by an arbitrary vector bundle
$\mathcal{E}\rightarrow M$, that we term a  {\emph{dynamics
bundle}}. The space of true physical trajectories is then
identified with zero locus of some section in the dynamics bundle:
$T\in \Gamma(\mathcal{E})$. (Relaxing the Helmholtz condition will
be the next step, addressed after relation (\ref{TV})). If $e^a$
is a local frame of sections of $\mathcal{E}$ over a trivializing
coordinate chart $U\in M$, so that $T=T_a(x)e^a$, then instead of
(\ref{1}) we get the equations of motion in the form:
\begin{equation}\label{T}
    T_a(x)=0 \, .
\end{equation}
Notice that we do not assume that $\dim T^\ast M=\dim
\mathcal{E}$, so the ``number'' of equations is allowed to be less
or greater than $\dim M$. (Of course, in the infinite dimensional
context, the notion of dimension needs clarification. An
appropriate definition can be done, for example, in the case of
local theories, i.e. when (\ref{T}) is a system of PDE's.)

The question arises what might be an analogue for the
integrability condition (\ref{dT}) that can ensure the existence
of a local action for the equations (\ref{T}). To answer this
question, let us first consider the case when $\dim
\mathcal{E}=\dim T^\ast M$. In this case, the answer is given by
existence of a vector bundle isomorphism $\Lambda:
\mathcal{E}\rightarrow T^\ast M$  such that $T'=\Lambda(T)$ is a
closed 1-form. In terms of local coordinates this reads
\begin{equation}\label{Lambda}
T'_i(x)=\Lambda^a_i(x)T_a(x) \, ,
\end{equation}
where $\Lambda^a_i(x)$ is a non-degenerate matrix. The equations
$T'=0$ are obviously equivalent to Eqs.(\ref{T}) in the sense that
both have the same solutions. Checking the closedness condition
for the 1-form $T'$ leads to the following relations:
\begin{equation}\label{intcon}
    dT'=0\quad \Leftrightarrow\quad dT_a = \Lambda^bC_{ab}^dT_d+G_{ab}\Lambda^b\,.
\end{equation}
Here we consider $T_a$ and $\Lambda^a=\Lambda^a_idx^i$ as a
collection of $0$- and $1$-forms defined on a coordinate chart $U$
and labelled by index $a$. The structure functions $G_{ab}$ and
$C_{ab}^c $ are, respectively,  symmetric and antisymmetric in
indices $a,b$.  In particular, the functions $C_{ab}^c$ enter to
the Maurer-Cartan equation for the basis $1$-forms $\Lambda^a$:
\begin{equation}\label{MC}
    d\Lambda^a=C^a_{bc}\Lambda^b\wedge\Lambda^c\,.
\end{equation}
It is the question of finding the ``integrating multiplier''
$\Lambda$ or investigating obstructions to its existence that the
inverse problem of variational calculus deals with. As soon as
$\Lambda$ is known, one can define a local action $S(x)$ such that
$dS=T_a \Lambda^a \equiv T'$. Having the local action at hands,
one can perform a quasi-classical path integral quantization in a
vicinity of any classical solution.

Since $\Lambda: \mathcal{E} \rightarrow T^\ast M$ is an
isomorphism of vector bundles, there is the inverse map
$V=\Lambda^{-1}:\mathcal{E}^\ast \rightarrow TM$ defining (and
defined by) a section $V=V_a^i(x) e^a\otimes\partial_i\in \Gamma(
\mathcal{E}\otimes TM)$. The integrability condition
(\ref{intcon}) is then  equivalent to the following relation in
terms of $V_a$ and $T_a$:
\begin{equation}\label{TV}
V_a^i\partial_iT_b-V_b^i\partial_iT_a=C_{ab}^dT_d\,.
\end{equation}
Now one may forget about $\Lambda$, taking the last relation as a
definition of the integrability condition, valid for an arbitrary
vector bundle $\mathcal{E}\rightarrow M$. In doing so, one has no
need to require the homomorphism $V: \mathcal{E}\rightarrow TM$ to
be of constant rank over $M$. It is the map $V$ subject to
relations (\ref{TV}) which we shall call the \emph{Lagrange
anchor}. The general and precise definition of the Lagrange anchor
is given in the next section. When the equations of motion are
defined by a 1-form on $M$, i.e $\mathcal{E}=TM$, the relation
(\ref{TV}) still remains much less restrictive for the dynamics
than the Helmholtz condition (\ref{intcon}), as the anchor
$V^i_j(x)$ satisfying (\ref{TV}), is not required to be
invertible. The usual Lagrangian dynamics (\ref{1}) corresponds to
the special case where $\mathcal{E}=TM$ and
$V^j_i=\delta_i^j=\Lambda_i^j$.

For the  general Lagrange anchor  (\ref{TV}), no integrating
multiplier $\Lambda = V^{-1}$ is required to exist. So, in
general, existence of the Lagrange anchor (\ref{TV}) does not mean
existence of a local action $S_U$ in the vicinity $U$ of given
solution $x_0$. Nonetheless, if $x_0$ is a \textit{regular} point,
in the sense that the ranks of matrices $(V^i_a)$ and $(\partial_i
T_a)$ are constant over $U$, the following statement holds true:
\begin{proposition}\label{LA}
Given a pair of sections $(T,V)$ satisfying (\ref{TV}), then for
any regular solution $x_0\in M$ of (\ref{T}) one can find a
coordinate system $(y^1,...,y^m, z^1,...,z^k)$ centered at $x_0$
together with  a set of local functions $S(y)$,
$E^1(y),...,E^k(y)$ such that equations $T_a(y,z)=0$ are
equivalent to
$$
\frac{\partial S(y)}{\partial y^{I}}=0\,,\qquad
    z^J = E^J(y)\,,
$$
where $k=\mathrm{rank}(\partial_iT_a(x_0)) -
    \mathrm{rank}(G_{ab}(x_0))$ and $G_{ab}=
    V_a^i\partial_iT_b$.
\end{proposition}
We prove this  proposition in Sect. 4.6. It is natural to call the
function $S(y)$, depending on a part of degrees of freedom, a
\textit{partial action}. On the surface defined by the equations
for $z$'s\footnote{In fact, these equations are not required to be
explicitly solved w.r.t. $z$'s, just a unique existence is
required for the solution with appropriate initial data.}, the
dynamics become Lagrangian for the other degrees of freedom, more
or less in the same sense as the Hamiltonian dynamics become
symplectic upon reduction to a symplectic leaf of a regular
Poisson structure. It should be noted  that the path integral
quantization method developed in this paper does not require any
special coordinate system. Moreover, the method is insensitive to
the rank of the matrix $V_a^i(x)$ and remains applicable to the
cases where no (partial) action can exist.

Let $\Sigma$ denote the set of all solutions, i.e. $\Sigma=\{p\in
M|T_a(p)=0\}$. Equations of motion (\ref{T}) are called
\textit{dependent} if there is a vector bundle
$\mathcal{E}_1\rightarrow M$ and a bundle homomorphism $Z:
\mathcal{E}\rightarrow \mathcal{E}_1$ such that $T\in \mathrm{ker}
Z$ and $\mathrm{rank}(Z|_\Sigma)\neq 0$. In terms of local
coordinates this means
\begin{equation}\label{ZT}
    Z^a_AT_a\equiv 0\,,
\end{equation}
where the section $Z=Z^a_Ae_a\otimes e^A\in
\Gamma(\mathcal{E}^\ast\otimes \mathcal{E}_1)$ does not vanish on
$\Sigma$ identically.

Equations of motion (\ref{T}) are said to be \textit{gauge
invariant} if there exists a vector bundle
$\mathcal{E}_{-1}\rightarrow M$ together with a bundle
homomorphism $R: \mathcal{E}_{-1}\rightarrow TM$ such that the
corresponding section $R=R^i_\alpha e^\alpha\otimes\partial_i\in
\Gamma(\mathcal{E}^\ast_{-1}\otimes TM)$ does not vanish on
$\Sigma$ identically and
\begin{equation}\label{RT}
    R_\alpha^i\partial_iT_a|_\Sigma=0\,.
\end{equation}

For an ordinary gauge theory with action $S$ and  gauge symmetry
generators $R_\alpha^i\partial_i$, Rels. (\ref{ZT}), (\ref{RT})
take the form
\begin{equation}\label{a}
    R_\alpha^i\partial_iS=0\,, \qquad
    R_\alpha^i\partial_i\partial_jS|_{dS=0}=0\,.
\end{equation}
As is seen, in the ordinary Lagrangian gauge theory, the role of
$R$'s is two-fold: the same $R$'s generate the gauge symmetries of
the equations of motion (\ref{RT}) and describe the functional
dependence (the Noether identities) between them (\ref{ZT}). If
the equations follow from the action principle, the gauge symmetry
means dependence of equations of motion, and vice versa. In the
general (non-Lagrangian) case, the generators $R$'s and $Z$'s may
be completely independent from each other. In particular, it is
possible to have dependent but not gauge invariant equations of
motion and vice versa.

Rels. (\ref{ZT}), (\ref{RT}) considered independently from
(\ref{TV}) define a gauge algebra structure irrespectively to
existence of the Lagrangian or Hamiltonian formalism. The BRST
imbedding for such a generic gauge algebra was systematically
described in Ref.\cite{LS1} along the usual lines of the BRST
theory\footnote{ Also in Ref.\cite{LS1}, the stress was made on
consistent combining of the gauge algebra  relations (\ref{ZT},
\ref{RT}) and the (weak) Poisson structure. That was aimed at
deformation quantization of the generic gauge systems (not
necessarily having Poisson structure on $M$) through constructing
a star-product which is associative only for the on-shell gauge
invariants, not for all functions on $M$. Examples of the
non-Lagrangian and/or non-Hamiltonian models where the BRST
description is an efficient tool for studying the classical
dynamics, can be found, e.g., in Refs. \cite{BGST}, \cite{LS2}.}
\cite{HT}. Examining the compatibility conditions between the
defining relations  for the gauge algebra (\ref{ZT}, \ref{RT}) and
the Lagrange structure (\ref{TV}), one can find rich algebraic and
geometric structures that are systematically studied in the
following sections. These are the structures which provide the
possibility to (path-integral) quantize the system even if its
classical dynamics does not admit action principle.

\section{Regular Lagrange structure and
$S_\infty$-algebra}\label{rls}

To summarize the previous discussion in a more formal way, a
classical system is specified by a vector bundle
$\mathcal{E}\rightarrow M$ over the space of trajectories $M$, and
a section $T\in \Gamma(\mathcal{E})$ playing the role of equations
of motion. The space of true trajectories of the system is then
identified with zero locus $\Sigma$ of $T$: $\Sigma = \{x \in M |
T(x)=0 \}$. In what follows we refer to $\Sigma$ as a
\textit{shell}.

\subsection{Lagrange structure}\label{RC} By a \emph{Lagrange structure}\emph{}
for a classical system $(\mathcal{E}, T)$ we understand
$\mathbb{R}$-linear map $d_\mathcal{E}: \Gamma(\wedge^n
\mathcal{E})\rightarrow \Gamma(\wedge^{n+1}\mathcal{E})$ obeying
conditions:
\begin{enumerate}
    \item [(i)]  \,  $d_\mathcal{E} T=0$\,,
      \item [(ii)] \,  $d_\mathcal{E}$ is a derivation of degree
      1, i.e.
$$
 d_{\mathcal{E}}(A\wedge B)=d_\mathcal{E}A\wedge B +
(-1)^{n}A\wedge d_\mathcal{E} B\,,\qquad \forall A\in
\Gamma(\wedge^n \mathcal{E}), \;\forall B\in \Gamma(\wedge^\bullet
\mathcal{E})\,.
$$
\end{enumerate}
Here we identify $\Gamma(\wedge^0\mathcal{E})$ with $C^\infty(M)$.

\vspace{2mm} Due to the Leibnitz identity (ii),
in each trivializing chart $U\subset M$ the operator
$d_\mathcal{E}$ is completely specified by its action on
coordinate functions $x^i$ and basis sections $e^a$ of
$\mathcal{E}|_U$:
\begin{equation}\label{dE}
    d_\mathcal{E} x^i=V^i_a(x) e^a\,,\qquad d_\mathcal{E} e^a=-\frac12C_{bc}^a(x)e^b\wedge
    e^c\,.
\end{equation}
Applying $d_\mathcal{E}$ to the section $T=T_ae^a$, one can see
that the property (i) reproduces the integrability condition
(\ref{TV}):
$$0=d_\mathcal{E} T =\frac12(V^i_a\partial_i T_b -V_b^i\partial_iT_a-
C_{ab}^cT_c )e^a\wedge e^b\,.$$ The first relation from (\ref{dE})
means also that $d_\mathcal{E}$ defines a bundle homomorphism $V:
\mathcal{E}^\ast\rightarrow TM$.

In the particular case where ${d_\mathcal{E}}^2=0$, $T$ is nothing
but a closed $1$-$\mathcal{E}$-form associated to the Lie
algebroid with the anchor $V$. Although for the general Lagrange
structure (i)-(ii), the differential $d_\mathcal{E}$ is not
required to be nilpotent, we call $V$ the \emph{Lagrange anchor}.
For the Lagrange anchor, the requirement of identical nilpotency
is replaced by a relaxed condition ${d_\mathcal{E}}^2 T = 0$
following from (i). This weaker requirement can have further
off-shell consequences that are derived in the next section under
certain regularity conditions on $T$.

\subsection{Regularity conditions}\label{regcon}
 Let $(\mathcal{E},T,d_\mathcal{E})$ be a Lagrange structure with shell  $\Sigma$.
 The Lagrange structure is said to be \textit{regular of type}
$(m,n)$ if  $\Sigma\neq\emptyset$ and there exists a finite chain
of vector bundles $\mathcal{E}_k\rightarrow M$ together with
$M$-bundle homomorphisms
\begin{equation}\label{exseq}
   0\rightarrow \mathcal{E}_{-m}{\rightarrow}\cdots\rightarrow
   \mathcal{E}_{-1}\stackrel{R}{\longrightarrow}TM\stackrel{J}{\longrightarrow}\mathcal{E}\stackrel{Z}
   {\longrightarrow}\mathcal{E}_{1}{\rightarrow}
    \cdots{\rightarrow}\mathcal{E}_{n}\rightarrow 0\,,
\end{equation}
such that
\begin{enumerate}
    \item [(a)]the map $J$ is defined by section $\nabla T\in \Gamma(T^\ast M\otimes
    \mathcal{E})$, where $\nabla$ is any connection on $\mathcal{E}$;
    \item [(b)]there is a neighbourhood $U\subset M$ of $\Sigma$ such
    that all the homomorphisms (\ref{exseq}) have constant ranks over $U$;
    \item [(c)]upon restriction to $\Sigma$, the chain (\ref{exseq}) makes an exact sequence.
\end{enumerate}

Several remarks are in order concerning this definition.

\textit{Remark 1.} The regularity condition ensures that
$\Sigma\subset M$ is a smooth submanifold.

\textit{Remark 2.} When exist, the homomorphisms (\ref{exseq}) are
not unique off shell. Thinking of these homomorphisms as sections
of the corresponding  vector bundles,
\begin{equation}\label{RTZ}
R=R^i_\alpha e^\alpha\otimes\partial_i\,,\quad J=\nabla_iT_a
dx^i\otimes e^a\,,\quad Z=Z_A^ae_a\otimes e^A\,, \quad ...,
\end{equation}
one can add to them any sections vanishing on $\Sigma$, leaving
the properties (a)-(c) unaffected. In particular, making a shift
\begin{equation}
Z\rightarrow Z+T_aW_A^{ab}e^A\otimes e_b\,,
\end{equation}
if necessary, we can always choose $Z$ in such a way  that $T \in
\ker Z$, cf. (\ref{ZT}).

\textit{Remark 3.} In the definition above one can pass from  the
chain (\ref{exseq}) to the transpose one by replacing each vector
bundle with its dual and inverting all the arrows. The transpose
chain meets the same conditions (a)-(c) as the original one.

\textit{Remark 4.}  The condition (c) means that $R$'s and $Z$'s,
defining the chain links in (\ref{exseq}), are to be understood as
the generators of gauge symmetry (\ref{RT}) and the generators of
Noether identities (\ref{ZT}). Having in mind this interpretation,
we term $\mathcal{E}_{-1}$ and $\mathcal{E}_1$ the \textit{gauge
algebra bundle} and the \textit{Noether identity bundle},
respectively. Accordingly, $\mathcal{E}$ is referred to as the
\textit{dynamics bundle }.

\vspace{2mm}

In this paper we deal mostly  with quantization of regular
$(1,1)$-type Lagrange structures associated to the four-term
sequences
\begin{equation}\label{4t-seq}
    0\rightarrow \mathcal{F}\stackrel{R}{\longrightarrow}TM
    \stackrel{J}{\longrightarrow}\mathcal{E}\stackrel{Z}
    {\longrightarrow}\mathcal{G}\rightarrow
    0\, .
\end{equation}
In other words,  we consider a set of gauge invariant and linearly
dependent equations of motion (\ref{T}), (\ref{ZT}), (\ref{RT})
with the generators of gauge symmetry $R_\alpha$ and Noether
identities $Z_A$  chosen in a linearly independent way. In the
ordinary Lagrangian gauge theory the dynamics bundle coincides
with the cotangent bundle ($\mathcal{E}=T^\ast M$), the Noether
identity bundle coincides with the gauge algebra bundle
($\mathcal{E}_{-1}=\mathcal{E}_1$) and the generators of gauge
symmetry coincide with the generators of Noether identities
($R=Z$). For the general system of type $(1,1)$, any of these
coincidences should not necessarily occur, e.g.: the gauge algebra
bundle $\mathcal{E}_{-1}$ and the bundle of Noether identities
$\mathcal{E}_1$ can be different even by dimension. In Sect.
\ref{Max} we give an example of quantizing the dynamical system of
type (0,1) which is not Lagrangian, although it has a nontrivial
Lagrange anchor. This means the theory has dependent equations of
motion having no gauge symmetry.

\subsection{Completeness}
A regular Lagrange structure $(\mathcal{E},T,d_\mathcal{E})$ is
called \textit{complete} if
\begin{equation}\label{}
    TM=\mathrm{Im}V\cup\mathrm{Im}R\,,
\end{equation}
where $V: \mathcal{E}^\ast\rightarrow TM$ is the Lagrange anchor
corresponding to $d_\mathcal{E}$, and $R$ is determined by
(\ref{exseq}). In other words, the completeness means that the
tangent bundle is spanned by the Lagrange anchor and the gauge
symmetry generators. It is easy to find that the number $m$ of
Lagrangian equations in Proposition \ref{LA} is equal to the rank
of the matrix $(R^i_\alpha, V^i_a)$. Hence, for a complete
Lagrange structure, all the equations of motion turn out  to be
(locally) Lagrangian. In this paper we consider regular Lagrange
structures that are not necessarily complete.

\subsection{Physical observables} Given a regular Lagrange
structure $(\mathcal{E},T,d_{\mathcal{E}})$ with the shell
$\Sigma$, we say that $f\in C^{\infty}(M)$ is a \textit{trivial}
function  if it vanishes on shell, i.e. $f|_\Sigma=0$. The
subspace of trivial functions is denoted by
$C^\infty(M)^{\mathrm{triv}}$. It follows from the regularity
conditions that $$f \in C^\infty(M)^{\mathrm{triv}}
\quad\Leftrightarrow\quad f =K^aT_a\,,$$ for some $K\in
\Gamma(\mathcal{E}^\ast)$.  Function $f\in C^{\infty}(M)$ is said
to be \textit{invariant} if for any section $\varepsilon=
\varepsilon^\alpha e_\alpha \in \Gamma(\mathcal{E}_{-1})$ there
exist a section $F\in \Gamma(\mathcal{E}^\ast\otimes
\mathcal{E}^\ast_{-1})$ such that
\begin{equation}\label{inv}
\varepsilon^\alpha R_\alpha^i\partial_i f =\varepsilon^\alpha
F_\alpha^aT_a \in C^\infty(M)^{\mathrm{triv}}\,,
\end{equation}
Here  $\{R_\alpha\}$ is an (over)complete basis of gauge
generators (\ref{RT}). Again, in view of the regularity conditions
one can rewrite (\ref{inv}) in a more compact way:
$R_\alpha^i\partial_i f|_{\Sigma}=0$.  The subspace of invariant
functions is denoted by $C^\infty(M)^{\mathrm{inv}}$. In view of
condition (\ref{RT}) the trivial functions are automatically
invariant, so we can write
$$C^\infty(M)^{\mathrm{triv}}
\subset C^\infty(M)^{\mathrm{inv}}\subset C^{\infty}(M)\,.$$ Two
invariant functions $f_1, f_2 \in C^\infty(M)^{\mathrm{inv}}$ are
considered as equivalent if they coincide on shell,
\begin{equation}\label{sim}
f_1 \sim f_2 \; \Leftrightarrow\; f_1-f_2 = K^aT_a \in
C^\infty(M)^{\mathrm{triv}}\,.
\end{equation}
The space of physical observables $\mathcal{P}$ is now defined as
the quotient of the space of invariant functions by the space of
trivial ones,
\begin{equation}\label{Ph}
\mathcal{P}=C^\infty(M)^{\mathrm{inv}}/C^\infty(M)^{\mathrm{triv}}
\end{equation}
Let us recall that $M$ is an infinite dimensional space of
trajectories, and $T_a=0$ are the differential equations whose
solutions are parametrized by initial data. The initial data
transformed into each other by the gauge symmetry transformations
are considered as equivalent. The space of inequivalent initial
data can then be  understood as a physical phase space $M_{phys}$.
In the case where $M_{{phys}}$ happen to be a smooth Hausdorf
manifold we have an equivalent definition of $\mathcal{P}$ as the
space of smooth functions on $M_{{phys}}$, i.e.
$\mathcal{P}=C^{\infty}(M_{\mathrm{phys}})$.

\subsection{$S_\infty$ - algebras}
An $S_\infty$-algebra ($S$ for Schouten) is a
$\mathbb{Z}_2$-graded, supercommutative and associative algebra
$A$ endowed with a sequence of odd linear maps $S_n: A^{\otimes
n}\rightarrow A$ such that
\begin{enumerate}
    \item[(i)]
    $S_n(...,a_k,a_{k+1},...)=(-1)^{\epsilon(a_k)\epsilon(a_{k+1})}S_n(...,a_{k+1},a_k,...)$,\\
    $\epsilon(a)$ being the parity of a homogeneous element $a\in
    A$.
    \item [(ii)]$a \mapsto S_n(a_1,...,a_{n-1},a)$ is a derivation
    of $A$ of the parity $1+\sum_{k=1}^{n-1}\epsilon(a_k)\; (\mathrm{mod}\;
    2)$.
    \item [(iii)]For all $n\geq 0$,
$$
\sum_{k+l=n}\sum_{(k,l)-\mathrm{shufle}} (-1)^\epsilon S_{l+1}
(S_k(a_{\sigma
(1)},...,a_{\sigma(k)}),a_{\sigma(k+1)},...,a_{\sigma(k+l)})=0\,,
$$
where $(-1)^\epsilon $ is the natural sign prescribed by the sign
rule for a permutation of homogeneous elements $a_1,...,a_n\in A$.
\end{enumerate}
Recall that a $(k,l)$-shuffle is a permutation of indices
$1,2,...,k+l$ satisfying $\sigma(1)<\cdots < \sigma(k)$ and
$\sigma(k+1)<\cdots<\sigma(k+l)$.

When $S_0=0$ we say about a \textit{flat} $S_\infty$-algebra. In
this case $S_1: A\rightarrow A$ is a nilpotent differential, and
$S_2$ induces an odd Poisson structure on corresponding
cohomology. An odd Poisson algebra can thus be regarded as
$S_\infty$-algebra with bracket $S_2:A\otimes A\rightarrow A$ and
all other $S_k=0$. In fact, properties (i) and (iii) characterize
$L_\infty$-algebras. See \cite{Vo} for recent discussion of
$S_\infty$-algebras.

It turns out that any regular Lagrange structure of type $(m,n)$
gives rise to an $S_\infty$-algebra structure on the
supercommutative algebra of sections

\begin{equation}\label{A}
A= \Gamma\big(\wedge^\bullet\mathcal{E}\otimes\bigotimes_{k=1}^m
S^\bullet (\Pi^k \mathcal{E}_{-k})\otimes\bigotimes_{l=1}^n
    S^\bullet(\Pi^{l+1}
    \mathcal{E}_l)\big)\,.
\end{equation}
Here $S^\bullet $ stands for symmetric tensor powers (in the
$\mathbb{Z}_2$-graded sense) and $\Pi$ denotes the parity
reversion operation, i.e. $\Pi \mathcal{E}$ is a vector bundle
over $M$ whose fibers are odd linear spaces. By definition,
$\Pi^2=\mathrm{id}$ and $S^\bullet(\Pi \mathcal{E})=\wedge^\bullet
\mathcal{E}$.

In the next section, applying the machinery of the BRST theory, we
give an explicit description for $S_\infty$-algebras associated
with $(1,1)$-type Lagrange structures. Extension to the
$(m,n)$-type Lagrange structures is straightforward.

\section{BRST imbedding}

\subsection{Ambient Poisson supermanifold} Let $(\mathcal{E},T,d_\mathcal{E})$
 be a regular Lagrange structure of type $(1,1)$ corresponding to the four-term
sequence (\ref{4t-seq}). Following the general line of ideas of
BRST theory, we have to realize the original space of trajectories
$M$ as a body of an appropriate $\mathbb{Z}$-graded supermanifold
$\mathcal{N}$. Let us choose $\mathcal{N}$ to be the total space
of the vector bundle
\begin{equation}\label{N}
\mathcal{N}=\Pi (\mathcal{F}\oplus \mathcal{F}^\ast)\oplus T^\ast
M \oplus \Pi(\mathcal{E}\oplus \mathcal{E}^\ast)\oplus
(\mathcal{G}\oplus \mathcal{G}^\ast) \, ,
\end{equation}
where $\mathcal{F,E}$ and $\mathcal{G}$ are the bundles of gauge
algebra, dynamical equations and the Noether identities
respectively (\ref{4t-seq}), see Remark 4 of Sect. \ref{regcon}.
The base $M$ is canonically imbedded into $\mathcal{N}$ as zero
section. Besides the Grassman parity, the fibers of each direct
summand in (\ref{N}) are endowed with an additional
$\mathbb{Z}$-grading, called the \textit{ghost number}. For
simplicity, to avoid cumbersome sign factors, we assume the base
$M$ to be an ordinary (even) manifold, that corresponds to the
case of mechanical systems without fermionic degrees of freedom.
Then the Grassman parity of the fibers is correlated to
$\mathbb{Z}$-grading in a simple way: the even coordinates have
even ghost numbers, while the odd coordinates  have odd ghost
numbers.  We also equip $\mathcal{N}$ with a pair of auxiliary
$\mathbb{N}$-gradings called the \textit{momentum-} and
\textit{resolution degrees } ($m$- and $r$-degrees, for short)
that will be used later for proving the existence theorem for the
BRST charge. The information about the gradings of local
coordinates is arranged in the table:
$$
\begin{tabular}{|l|c|c|c|c|c|c|c|c|}
  \hline

base and fibers   &$M$ & $T^\ast M$ & $\mathcal{F}
$&$\mathcal{F}^\ast$ &$\mathcal{E}$&$
\mathcal{E}^\ast$&$\mathcal{G}$& $\mathcal{G}^\ast$\\
\hline

local coordinates           &\,$x^i$ & $\bar x_j$&\, $c^\alpha
$\,& \,$\bar c_{\beta}$\, &\, $\eta_a$ & \,$\bar
\eta^b$&\,$\xi_A$\,&\,$\overset{}{\bar
           \xi}{}^B$\\
  \hline
  $\epsilon$=Grassman's parity            & 0 & 0 & 1 & 1 & 1 & 1 & 0 & 0\\
  \hline
  $\gh$ = ghost number                   & 0 & 0 & 1 & -1 & -1 & 1 & -2&2\\
  \hline
  $\mathrm{Deg}$ = momentum degree      & 0 & 1 & 0 & 1 & 0 & 1&0&1 \\
  \hline
  $\deg$ = resolution degree               & 0 & 1 & 0 & 2 & 1 & 0& 2& 0\\
  \hline
\end{tabular}
$$
\begin{center}
{Table 1}
\end{center}
 Splitting all the local
coordinates into the ``position coordinates''
$\varphi^I=(x^i,c^\alpha, \eta_a, \xi_A)$ and ``momenta'' $\bar
\varphi_J=(\bar x_i, \bar c_\alpha, \bar \eta^a, \bar\xi^A)$, we
can write
\begin{equation}\label{}
\begin{array}{ll}
    \gh(\bar\varphi_I)=-\gh(\varphi^I)\,,&\qquad
    \epsilon(\bar\varphi_I)=\epsilon(\varphi^I)\,,\\[3mm]
    \mathrm{Deg}
    (\bar\varphi_I)=1\,,&\qquad \mathrm{Deg}(\varphi^I)=0\,.
    \end{array}
\end{equation}
Though the submanifold $T^\ast M\subset \mathcal{N}$ is an
ordinary manifold, we do not include the fibers of $T^\ast M$ into
the body of $\mathbb{Z}\times\mathbb{Z}_2$-graded manifold $N$ and
treat $\bar x$'s as formal variables.

Fixing a linear connection
$\nabla=\nabla_{\mathcal{F}}\oplus\nabla_\mathcal{E}\oplus
\nabla_\mathcal{G}$ on $\mathcal{F}\oplus \mathcal{E}\oplus
\mathcal{G}$,  we endow $\mathcal{N}$ with the exact symplectic
structure $\omega =d\Lambda$, where
\begin{equation}\label{Theta}
\begin{array}{c}    \Lambda =\bar x_idx^i + \bar
    c_{\alpha}\nabla c^\alpha +\bar\eta^a \nabla \eta_a + \bar\xi^A\nabla
    \xi_A\,,\\[3mm]
\nabla c^\alpha=dc^\alpha+dx^i\Gamma_{i \beta}^\alpha c^\beta\,,
\end{array}
\end{equation}
and similar expressions are assumed for covariant differentials of
$\eta$'s and $\xi$'s. The corresponding Poisson brackets of local
coordinates read
\begin{equation}\label{brcov}
\begin{array}{lll}
 \{\bar\eta^b,\eta_a\}=\delta_a^b \,,\quad&  \{\bar x_i, \eta_a\}=\Gamma_{i a}^b
 \eta_b
 \,,\quad&
 \{\bar x_i,\bar\eta^b\}=-\Gamma_{ia}^b\bar\eta^a\,, \\[3mm]
 \{\bar c_\alpha,c^\beta\}=\delta^\beta_\alpha\,,\quad& \{\bar x_i,c^\alpha \}=\Gamma_{i\beta}^\alpha c^\beta\,
 \,,\quad&
 \{\bar x_i,\bar c_\beta\}= -\Gamma_{i\beta}^\alpha \bar c_\alpha\,,
  \\[3mm]
  \{\bar\xi^A,\xi_B\}=\delta_B^A\,,\quad& \{\bar
  x_i,\xi_A\}=\Gamma_{iA}^B\xi_B\,,\quad& \{\bar
  x_i,\bar \xi^A\}=-\Gamma_{iB}^A\bar\xi^B\,,
  \end{array}
\end{equation}
$$
\begin{array}{ll} \quad\{\bar x_i, x^j\}=\delta_i^j\,, \quad& \;\;\{\bar x_i,\bar
x_j\}=R_{ija}^b \bar\eta^a\eta_b+
  R_{ij\alpha}^\beta c^\alpha\bar
  c_\beta+R_{ijA}^B\bar\xi^A\xi_B\,,
  \end{array}
$$
and the other brackets vanish. The structure functions determining
the Poisson brackets of $\bar x_i$ and $\bar x_j$ are just
components of the curvature tensor of $\nabla$.

Clearly, the equations $\bar \varphi_I=0$ define the Lagrangian
submanifold
\begin{equation}\label{AntiP}
\mathcal{L}=\Pi (\mathcal{F}\oplus  \mathcal{E})\oplus \mathcal{G}
\subset \mathcal{N}\,,
\end{equation}
and the supercommutative algebra of functions
$C^\infty(\mathcal{L})$ is naturally isomorphic to the algebra
(\ref{A}) with $m=n=1$.

\subsection{BRST charge} It turns out that all the ingredients
of the Lagrange structure can be naturally interpreted as
coefficients of expansion in the fiber coordinates  of a single
function $\Omega\in C^{\infty}(\mathcal{N})$, called the BRST
charge, such that
\begin{equation}\label{16}
\gh(\Omega)=1\,,\qquad \epsilon(\Omega)=1\,,\qquad
\mathrm{Deg}(\Omega)\geq 1\,.
\end{equation}
The relations defining the Lagrange structure are generated by and
are equivalent to the master equation
\begin{equation}\label{me}
    \{\Omega,\Omega\}=0\,.
\end{equation}
To get the desired interpretation let us first expand $\Omega$ in
the powers of momenta $\bar\varphi_I$:
\begin{equation}\label{m-exp}
\Omega =\sum_{n=1}^\infty \Omega_n\,,\qquad
\mathrm{Deg}(\Omega_n)=n\,.
\end{equation}
Substituting this expansion into the master equation (\ref{me})
and considering it in the first three orders in the momenta, we
get the following relations for $\Omega_1,\Omega_2$:
\begin{equation}\label{waps}
    \{\Omega_1,\Omega_1\}=0\,,\qquad
    \{\Omega_1,\Omega_2\}=0\,,\qquad
    \{\Omega_2,\Omega_2\}=2\{\Omega_1,\Omega_3\}\,.
\end{equation}
The first term $\Omega_1=\Omega^I(\varphi)\bar\varphi_I$ gives
rise to the odd, nilpotent vector field on  $\mathcal{L}$,
\begin{equation}\label{Q}
    Q\equiv\Omega^I(\varphi)\frac{\partial}{\partial \varphi^I} =
    T_a\frac{\partial}{\partial \eta_a}+c^\alpha
    R_\alpha^i\frac{\partial}{\partial x^i}+\eta_aZ^a_A\frac{\partial}{\partial
    \xi_A}+\cdots\,,
\end{equation}
carrying all the information about the classical system
$(\mathcal{E},T)$ itself. Evaluating the nilpotency condition
$Q^2=0$ at the lowest order in $r$-degree (see Table 1), one
immediately recovers Rels.(\ref{ZT}, \ref{RT}) characterizing
(\ref{T}) as a set of gauge invariant and linearly dependent
equations of motion, with $R$ and $Z$ being the generators of
gauge transformations and Noether identities,
respectively\footnote{As it has been already mentioned, these
generators, coinciding in  the standard Lagrangian case, can be
different in general, even by number.}.

Notice that the odd vector fields with zero square are known as
\textit{homological}. These are essentially equivalent to the
notion of $L_\infty$-\textit{algebras} or \textit{strong(ly)
homotopy Lie algebras} \cite{HiS}, \cite{SL}. In physical
literature, the homological vector fields usually appear  as
\textit{BRST-differentials} associated either with the BV master
action \cite{BV}, \cite{HT} or Hamiltonian BFV-BRST charge
\cite{BFV}, \cite{HT}.

The Poisson action of $\Omega_1$ makes $C^\infty (\mathcal{N})$ a
cochain complex with the nilpotent differential
\begin{equation}\label{D}
    \mathbb{D}A=\{\Omega_1,A\}\,,\qquad \forall A\in
    C^{\infty}(\mathcal{N})\,.
\end{equation}
 Let us denote by
$\mathcal{H}(\mathbb{D})=\bigoplus_{n=0}^\infty
\mathcal{H}^n(\mathbb{D})$ the corresponding cohomology group
graded by $m$-degree.

The Lagrange anchor $V: \mathcal{E}^\ast\rightarrow TM$,
associated to the Lagrange structure for the classical system
(\ref{Q}) is contained in the next term
\begin{equation}
\Omega_2=\Omega^{IJ}(\varphi)\bar\varphi_I\bar\varphi_J=\bar\eta^aV_a^i\bar
x_i+\cdots\,.
\end{equation}

Rels.(\ref{waps}) characterize $\Omega_2$ as a weak anti-Poisson
structure on $\mathcal{L}$ (\ref{AntiP}), i.e.
$\Omega_1$-invariant (= $\mathbb{D}$-closed) odd bivector
satisfying the Jacobi identity up to homotopy. The corresponding
weak antibracket reads
\begin{equation}\label{wabr}
(a,b)\equiv\frac12\{\{\Omega_2,a\},b\}\,,\qquad a,b \in
C^{\infty}(\mathcal{L})\,.
\end{equation}
Examining  the Jacobi identity for these brackets one finds
\begin{equation}\label{}
\begin{array}{c}
    (a,(b,c))+(-1)^{\epsilon(b)\epsilon(c)}((a,c),b)+
    (-1)^{\epsilon(a)(\epsilon(b)+\epsilon(c))}((b,c),a)=\\[3mm]
- S_3(
\mathbb{D}a,b,c)-(-1)^{\epsilon(a)\epsilon(b)}S_3(a,\mathbb{D}b,c)
-(-1)^{(\epsilon(a)+\epsilon(b))\epsilon(c)}S_3(a,b,\mathbb{D}c)\\[3mm]-
\mathbb{D}S_3(a,b,c)\,.
    \end{array}
\end{equation}
where we have introduced the following notation:
\begin{equation}\label{Sn}
    S_n(a_1,a_2,...,a_n)\equiv\frac1{n!}\{...\{\Omega_n,a_1\}a_2\},...,a_n\}\,,\qquad a_k\in
    C^{\infty}(\mathcal{L})\,.
\end{equation}
In particular,
\begin{equation}
\begin{array}{c}
S_0\equiv0\,,\quad S_1(a)=\mathbb{D}a=\Omega^I\partial_Ia\,, \quad
S_2(a,b)=(a,b)=(-1)^{\epsilon(a)\epsilon(I)}\Omega^{IJ}\partial_J
a\partial_Ib\,, \\[3mm]
S_3(a,b,c)=(-1)^{\epsilon(a)(\epsilon(I)+
\epsilon(J))+\epsilon(b)\epsilon(I)}\Omega_3^{IJK}\partial_Ka\partial_Jb\partial_Ic\,.
\end{array}
\end{equation}
As is seen, the weak antibracket (\ref{wabr}) induces the genuine
antibracket on the cohomology group $\mathcal{H}^0(\mathbb{D})$.

As was noticed in \cite{Vo},  Rel. (\ref{Sn}) defines a flat
$S_\infty$-structure on the supercommutative algebra
$C^\infty(\mathcal{L})$: By definition, each $S_n$ is a
graded-symmetric multi-differentiation of $C^\infty(\mathcal{L})$
and the generalized Jacobi identities for the collection of maps
$\{S_n\}$ follow from the master equation (\ref{me}) for the BRST
charge $\Omega$.

\subsection{The unique existence of the BRST charge}
In our treatment of the Lagrange structures the BRST charge
$\Omega$ is not given  \textit{a priory} - it arises as a solution
to the master equation (\ref{me}) with prescribed ``boundary
conditions''. By the boundary conditions we mean the starting row
of expansion of the BRST charge according to $r$-degree:
\begin{equation}\label{bcon}
\Omega =T_a\bar\eta^a+c^\alpha
R_\alpha^i\bar{x}_i+\eta_aZ^a_A\bar\xi^A +\bar\eta^aV^i_a\bar{x}_i
+\cdots \, ,
\end{equation}
where the dots stand for the terms more than quadratic in the
fiber coordinates  and/or $r$-degree $>1$. The structure functions
$T$'s, $R$'s, $Z$'s and $V$'s are naturally identified with the
equations of motions, gauge symmetry generators, generators of
Noether identities and the Lagrange anchor, respectively.

Given the boundary conditions (\ref{bcon}), the existence of a
solution to the master equations (\ref{me}) can be proved by the
standard tools of homological perturbation theory \cite{HT},
\cite{St}. Let us sketch this proof. Consider the following
expansion of the BRST charge according to the $r$-degree:
\begin{equation}\label{rd-exp}
    \Omega=\sum_{n=0}^\infty \Omega^{(n)}\,,\qquad
    \deg\left(\Omega^{(n)}\right)=n\,.
\end{equation}
In particular, the general expressions for the first two terms
read
\begin{equation}\label{exexp}
\Omega^{(0)}=T_a\bar\eta^a\,,\quad \Omega^{(1)}=c^\alpha
R_\alpha^i\bar x_i+\eta_aZ^a_A\bar\xi^A +\bar\eta^aV^i_a\bar x_i
+\bar\eta^a\eta_b U_{\alpha a}^bc^\alpha +\bar\eta^a\bar\eta^b
W_{ab}^d\eta_d\,.
\end{equation}
Substituting (\ref{rd-exp}) into the master equation (\ref{me})
gives a chain of equations of the form:
\begin{equation}\label{sn}
    \delta \Omega^{(n+1)}=K_n(\Omega^{(0)},...,\Omega^{(n)})
    \,,\qquad n=0,1,2,..\,,
\end{equation}
where
\begin{equation}\label{delta}
    \delta =T_a\frac{\partial}{\partial \eta_a}+\eta_aZ^a_A\frac{\partial}{\partial \xi_A}+\bar x_iR^i_\alpha\frac{\partial}{\partial\bar c_\alpha}
+\bar\eta^a\nabla_iT_a\frac{\partial}{\partial\bar
x_i}+\bar\eta^a\eta_b U^b_{\alpha a}\frac{\partial}{\partial\bar
c_\alpha}
\end{equation}
is a nilpotent operator decreasing the $r$-degree by one,
\begin{equation}\label{d2}
\delta^2=0\,,\qquad\deg(\delta)=-1\,,
\end{equation}
and $K_n$ involves the brackets of the $\Omega$'s of lower order.
The nilpotency condition (\ref{d2}) is due to the following
relations (cf. Eqs.(\ref{ZT},\ref{RT})):
\begin{equation}\label{str-rel}
Z^a_AT_a=0\,,\qquad R_\alpha^i\nabla_iT_a=U_{\alpha a}^bT_b\,.
\end{equation}
Let $\mathcal{H}(\delta)=\bigoplus_{n=0}^\infty
\mathcal{H}_n(\delta)$ denote the corresponding cohomology group
graded by $r$-degree. It is not hard to see that the regularity
conditions of Sect.\ref{regcon}, which we assume satisfied for the
classical system under consideration, provide acyclicity of
$\delta$ in strictly positive $r$-degrees:
\begin{equation}\label{H}
\mathcal{H}_n(\delta)=0\, \quad \mathrm{for} \quad n>0\,.
\end{equation}
The proof of the last fact is quite standard (see e.g. \cite{HT})
and we leave it to the reader.

On the other hand, expanding the Jacobi identity
$\{\{\Omega,\Omega\},\Omega\}=0$ in terms of $r$-degree,  one can
deduce that the r.h.s. of the $(n+1)$-th equation (\ref{sn}) is
$\delta$-closed provided all the previous equations are satisfied.
Therefore, the only equation to check to ensure solvability of
(\ref{sn}) is
\begin{equation}\label{brel}
    \delta \Omega^{(1)} =K_0(\Omega^{(0)})\,,
\end{equation}
where $K_0(\Omega^{(0)})\equiv\{\Omega^{(0)},\Omega^{(0)}\}=0$.
Substituting the explicit expressions (\ref{exexp}), (\ref{delta})
into (\ref{brel}), we reproduce the basic relations
(\ref{str-rel}) as well as the integrability condition (\ref{TV})
with $C_{ab}^d=W_{ab}^d+V_a^i\Gamma_{ib}^d-V_b^i\Gamma_{ia}^d$,
where $W_{ab}^d$ is defined by (\ref{exexp}). Thus,
Eq.(\ref{brel}) generates all the defining relations for the
regular Lagrange structure $(\mathcal{E}, T,d_\mathcal{E})$ of
type $(1,1)$.

Resolving Eqs.(\ref{sn}) step-by-step, we are obviously free to
add to $n$-th order solution $\Omega^{(n)}$ any $\delta$-closed
(and hence exact) term. This ambiguity is not essential, however,
as it can always be absorbed by a canonical transformation of the
Poisson supermanifold $N$. The proof of the last fact is quite
standard (see e.g. \cite{HT}), and we omit it here.

\subsection{Exactness of Lagrange structure}\label{exact}
An important observation about the weak antibracket (\ref{wabr}),
encoding the Lagrange structure, is that $\Omega_2$ determines a
trivial $\mathbb{D}$-cocycle. In other words, for any $\Omega_2$
there is a function
\begin{equation}\label{GGG}
\begin{array}{c}
G = G^{ij}(x)\bar x_i\bar x_j+\cdots \,,\\[3mm]
\mathrm{Deg}(G)=2\,,\quad \epsilon(G)=0\,,\quad\mathrm{gh}(G)=0\,,
\end{array}
\end{equation}
such that $\Omega_2=\mathbb{D}G$. The last fact is a
straightforward consequence of a more general statement about
$\mathbb{D}$-cohomology.

\begin{proposition}\label{ex} Let $\mathcal{H}(\mathbb{D})=\bigoplus \mathcal{H}_n^m(\mathbb{D})$
be the group of $\mathbb{D}$-cohomology (\ref{D}), where the
numbers $m$ and $n$ refer to the $m$-degree and the ghost number
respectively, then $\mathcal{H}^m_n(\mathbb{D})=0$ for all $m>n$.
\end{proposition}
\begin{prf} We start with the following decomposition:
\begin{equation}\label{}
    \mathbb{D}=\delta +\Delta \,,
\end{equation}
where $\delta$ is given by (\ref{delta}) and
$$\qquad\deg(\delta)=-1\,,\quad
    \deg(\Delta)\geq 0\,.$$
Thus, $\mathbb{D}$ is a deformation of $\delta$ by terms of higher
$r$-degree and the statement follows immediately  from the
acyclicity of $\delta$ in positive $r$-degree. To make these
arguments more explicit let us introduce  the contracting homotopy
for $\delta$, i.e. an operator $\delta^\ast$ obeying the property
\begin{equation}\label{}
    (\delta \delta^\ast+\delta^\ast\delta)A=A\,,
\end{equation}
for all $A$ with $\deg(A)>0$. Using this operator, one can show
that  $\mathbb{D}$-cohomology is localized at zero $r$-degree.
Indeed, applying $\delta^\ast$ to both sides of $\mathbb{D}A=0$
yields
\begin{equation}\label{}
    NA=\mathbb{D}\delta^{\ast}A\,,\qquad N\equiv 1+(\delta^\ast \Delta+\Delta\delta^\ast)\,.
\end{equation}
Since $\deg(\delta^\ast \Delta+\Delta\delta^\ast)>0$, the operator
$N$ is invertible and commutes with $\mathbb{D}$. Thus $A =
\mathbb{D}(N^{-1}\delta^{\ast}A)$. To complete the proof it
remains to note that
\begin{equation}\label{ineq}
    \deg (A)\geq\mathrm{Deg}(A)-\mathrm{gh}(A)\,.
\end{equation}
The last inequality can be  verified just by comparing the lines
of Table 1.
\end{prf}

Applying now the inequality (\ref{ineq}) to $\Omega_2$, we get
$\deg(\Omega_2)\geq 1$, and hence $\Omega_2=\mathbb{D}G$. For a
reason that will become clear later on, we call the function $G$,
determined up to a $\mathbb{D}$-cocycle, a \textit{propagator}
associated to the weak anti-Poison structure $\Omega_2$.

The proposition above allows us to give another prove of the
existence theorem for the BRST charge $\Omega$. Namely, consider
the Hamiltonian flow $\phi_t$  generated by the propagator $G$.
Applying this flow to the symbol of homological vector field
$\Omega_1$, determined by Rels. (\ref{m-exp}, \ref{Q}), we get a
one-parameter family of formal functions $\Omega(t)\in
C^{\infty}(\mathcal{N})[[t]]$ related to each other by a formal
canonical transform. By definition,
\begin{equation}\label{omegat}
    \Omega(t)=\phi^\ast_t \Omega_1 = \sum_{n=0}^\infty \frac
    {t^n}{n!}\{G,\{G,\cdots,\{G,\Omega_1\}\cdots\}\,.
\end{equation}
Clearly, the function $\Omega(1)$ obeys the boundary condition
(\ref{bcon}) and solves the master equation (\ref{me}) whenever
$\Omega_1$ does. The problem thus reduces to solving the
nilpotency condition for the homological vector field (\ref{Q}).
Expanding $\Omega_1=\sum \Omega_1^{(n)}$ according to the
resolution degree, we get a chain of equations
\begin{equation}\label{qn}
    \delta \Omega_1^{(n+1)}=K_n(\Omega_1^{(0)},...,\Omega_1^{(n)})\,,
\end{equation}
which are closely analogous to Eqs. (\ref{sn}), though with
$\delta$ which is different from (\ref{delta}). Here
\begin{equation}\label{KT}
    \delta= T_a\frac{\partial}{\partial
    \eta_a}+\eta_aZ^a_A\frac{\partial}{\partial \xi_A}
\end{equation}
is the usual Koszul-Tate differential associated with the shell
$\Sigma$. The solvability of the system (\ref{qn}) can be easily
seen from the acyclicity  of $\delta$ in strictly positive
$r$-degree \cite{HT} and the invariance of $\Sigma$ under the
gauge transformations (\ref{RT}).

\subsection{Physical observables}\label{PO}
Upon the BRST imbedding, the physical observables of original
theory are usually identified with certain BRST cohomology in
ghost number zero. Let us show that the space of physical
observables $\mathcal{P}$, defined by (\ref{Ph}), is naturally
isomorphic to the subgroup $\mathcal{H}^0_0(\mathbb{D})\subset
\mathcal{H}(\mathbb{D})$ generated by the $\mathbb{D}$-cocycles of
the ghost and momentum degree zero. Substituting the general
expansion
\begin{equation}\label{}
    F=\sum_{n=0}^\infty F^{(n)}(\varphi)=f(x)+\eta_aF^a_\alpha(x) c^\alpha
    +\xi_A F^A_{\alpha\beta}(x)c^\alpha c^\beta+\cdots\,,\quad
    \mathrm{deg}(F^{(n)})=n\,,
\end{equation}
to the $\mathbb{D}$-closedness condition $\mathbb{D}F=0$ leads to
the sequence of equations
\begin{equation}\label{Fn}
    \delta F^{(n+1)}=B_n(F^{(0)},...,F^{(n)})\,,\qquad
    \mathrm{deg}(B_n)=n\,,
\end{equation}
where the $\delta$ is given by (\ref{KT}). The first equation of
this sequence reproduces the condition of on-shell invariance
(\ref{inv}). Proceeding by induction in $n$ and using the identity
$\mathbb{D}^2F=0$, one can see that the r.h.s. of the $n$-th
equation (\ref{Fn}) is $\delta$-closed, provided that Eq.
(\ref{inv}) is satisfied. Since the differential $\delta$ is
acyclic in positive $r$-degree, we conclude that (i) any invariant
function $f\in C(M)^{\mathrm{inv}}$ is lifted to a
$\mathbb{D}$-cocycle $F\in C^{\infty}(\mathcal{N})$ with
$\mathrm{gh}(F)=\mathrm{Deg}(F)=0$, and (ii) any two equivalent
(in the sense of (\ref{sim})) functions $f_1,f_2\in
C^{\infty}(M)^{\mathrm{inv}}$ determine the same class of
$\mathbb{D}$-cohomology upon the lift:
\begin{equation}\label{}
    F_1-F_2=\mathbb{D}K\,,\qquad K=K^a\eta_a+\cdots\,.
\end{equation}
This establishes an isomorphism between the space of physical
observables  $\mathcal{P}$ of the  original theory (\ref{Ph}) and
the  BRST cohomology group $\mathcal{H}^0_0(\mathbb{D})$.

\subsection{Lagrange structure from the Hamiltonian viewpoint.}
In the conventional BFV approach the BRST  charge arises as a tool
for quantizing first-class constrained Hamiltonian systems. Given
a Poisson manifold $P$, the first class constraints $\Theta_I\in
C^{\infty}(P)$ are defined as  an (overcomplete) basis in the
regular, Poisson-closed ideal of functions vanishing on a
coisotropic submanifold $C\subset P$. In a more general setting
\cite{LS2}, one can think of $\{\Theta_I\}$ as a section of a
(nontrivial) vector bundle $E \rightarrow P$, which intersects the
base $P$ at points of $C$. The standard BFV-BRST theory
\cite{BFV}, \cite{HT} corresponds to the case of a trivial vector
bundle $E$. According to the general prescriptions of BFV-BRST
theory, to each first class constraint, one has to associate a
pair of canonically conjugated ghost variables
$(\mathcal{C},\mathcal{P})$, extending thus the original Poisson
manifold $P$ to the supermanifold $\Pi (E\oplus E^\ast)$. In the
reducible case, i.e when the constraints $\Theta_I$ are
functionally dependent, the additional pairs of canonically
conjugated variables (ghosts-for-ghosts) must be introduced into
the scheme \cite{BFV}, \cite{HT}.

A glance at Table 1 is enough to see that the spectrum of ghost
numbers corresponds to Hamiltonian first-class constrained system
of a first-stage of reducibility. In order to make this
interpretation explicit let us combine the local coordinates with
ghost numbers $1$ and $-1$ into the ghost coordinates
$\mathcal{C}^I=(\bar \eta^a, c^\alpha)$ and ghost momenta
$\mathcal{P}_I=(\eta_a, \bar c_\alpha)$, respectively. In this
notation the BRST charge (\ref{bcon}) can be rewritten as
\begin{equation}\label{}
    \Omega = \mathcal{C}^I\Theta_I(x,\bar x)+ \mathcal{P}_I\Xi_A^I(x,\bar
    x)\xi^A+ \frac12 \mathcal{P}_K {U}^K_{IJ}(x,\bar
    x)\mathcal{C}^J\mathcal{C}^I + O(\mathcal{P}^2,\xi^2)\,,
\end{equation}
where the expansion coefficients  $\Theta_I
=(\tilde{T}_a,\tilde{R}_\alpha)$ and $\Xi^I_A=(\tilde{Z}_A^a,0)$,
playing the role of first class constraints and their
null-vectors, are given by the formal power series in $\bar x$'s
of the form
\begin{equation}\label{TRZ}
\begin{array}{l}
\tilde{T}_a(x,\bar x)=T_a(x)+V_a^i(x)\bar x_i+O(\bar x^2)\,,\\[3mm]
\tilde{R}_\alpha(x,\bar x)=R_\alpha^i(x)\bar x_i+O(\bar
x^2)\,,\\[3mm]
\tilde{Z}_A^a(x,\bar x)=Z_A^a(x)+O(\bar x)\,.
\end{array}
\end{equation}
At lowest orders in $\mathcal{C}$'s the master equation (\ref{me})
gives the standard involution relations for a set of reducible
first-class constraints
\begin{equation}\label{ThetaInv}
    \{\Theta_I,\Theta_J\}={U}_{IJ}^K\Theta_K\,,\qquad
    \Xi_A^I\Theta_I=0\,,
\end{equation}
w.r.t. the canonical Poisson bracket on $T^\ast M$. From the
regularity condition it readily follows that the number of the
independent  first class constraints $\Theta_I$ is equal to $\dim
M$. In physical terms, one can interpret this fact concluding that
the considered Hamiltonian system has no (local) physical degrees
of freedom. From the geometrical viewpoint this implies  that
equations $\Theta_I=0$ define a Lagrangian submanifold $L\subset
T^\ast M$; more accurately, $L$ is a formal Lagrangian submanifold
as we are not concerned with convergence of the formal series
(\ref{TRZ}).

The first class constraints $\Theta_I$ can also be regarded as a
formal deformation of those given by the leading terms of
expansions (\ref{TRZ}) in the ``direction'' of the Lagrange anchor
$V$. The involution relations for the ``initial'' constraints
$T_a(x)$ and $R_\alpha^i(x)\bar x_i$ readily follow from
nilpotency of the homological vector field (\ref{Q}).  The
integrability condition  (\ref{TV}) results then from the
requirement that the deformed constraints (\ref{TRZ}) have to be
the first class as well. From this standpoint,  the Lagrange
structure can be understood as an infinitesimal of deformation of
the Lagrangian submanifold $L_0 \subset T^\ast M$ defined by the
equations $T_a(x)=0$ and $ R^i_\alpha(x)\bar x_i=0$. As is shown
in the next section, any nonzero Lagrange anchor gives rise to
quantum fluctuations of physical observables. In other words, any
``classical'' deformation of $L_0\subset T^\ast M$ (in the
category of Lagrangian submanifolds in $T^\ast M$) results in a
quantum deformation upon path-integral quantization.

Actually, the results of Sect. \ref{exact} allows one to interpret
the series (\ref{TRZ}) as Tailor's expansion in $\bar x$'s of some
smooth functions $\Theta_I(x,\bar x)$, $\Xi_A^I(x,\bar x)$ defined
in a sufficiently small vicinity $U$ of any given point $p\in
T^\ast M$. This can be proved as follows. Since both the symbol
$\Omega_1$ of homological vector field (\ref{Q}) and the generator
$G$ of canonical transform (\ref{omegat}) are smooth function on
$\mathcal{N}$ (with polynomial dependence of fiber variables), one
can assert that for any $p\in \mathcal{N}$ there exists a
neighborhood $W$ together with $\varepsilon
>0$ such that $\Omega(t)=\phi_t\Omega_1$ is a smooth
function on $W$ for all $t\in [0,\varepsilon)$. Moreover,
shrinking the vicinity $W\subset \mathcal{N}$ along the fibers,
one can always choose  $\varepsilon>1$, so that
$$\left.\left.
\Theta_I(x,\bar x)\equiv \frac{\partial \Omega(1)}{\partial
\mathcal{C}^I}\right|_{\mathcal{C}=\xi=0}\,,\qquad \Xi_A^I(x,\bar
x)\equiv\frac{\partial^2 \Omega(1)}{\partial \xi^A\partial
\mathcal{P}_I}\right|_{\mathcal{C}=\xi=0}
$$
are smooth functions on an open subset $U=W\cap T^\ast M$.

Let $x_0\in \Sigma \subset M$ be a classical solution and let $U$
be a sufficiently small neighbourhood of $x_0$ in $T^\ast M$ for
which the equations $ \Theta_I(x,\bar x)=0 $ make sense, i.e.
determine a Lagrangian submanifold $L \subset U$. If
$\mathrm{rank}(V_a^i(x), R_\alpha^i(x))=m$ for all $x\in U\cap M$,
then one can split the position coordinates and momenta onto two
groups, $x^i=(y^I, z^J)$ and $\bar x_i=(\bar{y}_I,\bar z_J)$, such
that the index $I$ runs $m$ values, and the Lagrangian submanifold
$L\subset U$ is determined by the equations
\begin{equation}\label{GF}
\bar y_I=\frac{\partial \Psi(y,\bar z)} {\partial y^I}\,,\qquad
z^J=\frac{\partial \Psi(y,\bar z)}{\partial \bar z_J}\,,
\end{equation}
$\Psi$ being the generating function for $L$ of the first kind
\cite{W}. Indeed, due to the rank condition the equations
$\Theta_I(x,\bar x)=0$ can be explicitly resolved  w.r.t. $m$ of
$\dim M$ momenta $\bar x_i$. Then using the rest of the equations
one can express $\dim M-m$ position coordinates $x$'s in terms of
the other variables (as the total number of independent equations
equals  $\dim M$.)  Since  the resolved constraints are always in
the abelian involution, one finds immediately that the r.h.s. of
these constraints are to be given by the gradient of some function
$\Psi(y,\bar z)$.

Now we can use the local representation (\ref{GF}) for $L$ to
prove the announced Proposition \ref{LA}. Clearly, the classical
equations of motion $T_a(x)=0$ are equivalent to $\Theta_I(x,\bar
x)=0$ and $\bar x_i=0$. (Geometrically speaking, the shell
$\Sigma$ is given by the intersection $L\cap M$.) Setting in Eqs.
(\ref{GF}), $\bar x_i=0$ yields the following local representation
for the shell:
\begin{equation}\label{SE}
    \frac{\partial S(y)}{\partial y^I}=0\,,\qquad
    z^J=E^J(y)\,,
\end{equation}
where
\begin{equation}
S(y)\equiv\Psi(y,0)\,,\qquad E^J(y)\equiv\frac{\partial \Psi}
{\partial \bar z_J}(y,0)\,.
\end{equation}
A simple linear algebra shows that the number $k=\dim M-m$  of
non-Lagrangian equations of motion is defined by the formula of
Proposition \ref{LA}.

\section{Quantization}
The usual path-integral quantization deals with computing of
quantum averages for the physical observables (= function(al)s on
the space of trajectories $M$). The quantum average of an
observable $F$ is given by its integral over $M$ with the uniform
weight $e^{\frac i\hbar S}$, $S$ being the action functional of
the system. In the gauge invariant Lagrangian theory, this simple
rule is replaced by a more sophisticated BV scheme
\cite{BV},\cite{HT} realizing the same idea in the presence of
gauge invariance. If the original classical system admitted
consistent operator BFV-BRST quantization, the BV method can be
deduced \cite{BF} from the Hamiltonian BFV-BRST scheme. In
general, the relationship remains obscure between the BV
quantization and the deformation quantization of the (weak)
Poisson manifolds, so the BV method is presently viewed as an
\textit{ad hoc} postulate for the path-integral quantization.
Obviously, the BV scheme can not be directly applied to quantize
the systems having no action functional. In this section, we
extend the path-integral quantization method to include not
necessarily Lagrangian (gauge) systems.

Let us briefly outline the quantization algorithm we suggest. The
starting point is the BRST embedding for the Lagrange structure
presented in the previous section.  Upon this embedding, the
physical observables of the original theory are identified with
the BRST cohomology group $\mathcal{H}_0^0(\mathbb{D})$. As the
next step, making use of the AKSZ method \cite{AKSZ} in the form
of Ref. \cite{GD}, we construct a topological sigma-model related
to this BRST complex. Then we prove that the dynamics of this
topological sigma-model are equivalent to the original classical
theory; in so doing, the space of physical observables
$\mathcal{H}_0^0(\mathbb{D})$ is naturally identified with the
boundary observables of the topological sigma-model. The
topological sigma-model, being a Lagrangian theory in usual sense,
can be quantized by the standard BV prescription, that results in
quantizing the original theory which is not necessarily
Lagrangian. In a particular case of Lagrangian systems, the
sigma-model path-integral can be explicitly localized at the
boundary, where it precisely reproduces the BV answer for the
original Lagrangian theory.

\subsection{Topological sigma-model} Consider the $(1,1)$-dimensional
supermanifold $\mathcal{I}=\Pi TI$ with boundary associated to the
odd tangent bundle of the closed interval $I=[0,1]\subset
\mathbb{R}$. The ``points'' of $\mathcal{I}$ are parameterized by
one even coordinate $t\in [0,1]$ of ghost number $0$ and one odd
coordinate $\theta\in \Pi T_tI$ of ghost number $1$. We shall also
use the collective notation $z=(t,\theta)$.  By the boundary of
$\mathcal{I}$ we mean the two-point set $\partial
\mathcal{I}=\{z_0, z_1\}\subset \mathcal{I}$ constituted  by the
``end points'' $z_0=(0,0)$ and $z_1=(1,0)$ of the superinterval
$\mathcal{I}$. The canonical volume element on $\mathcal{I}$ is
given by $d^2z=dt d\theta$.

Consider now the superspace  $\mathcal{N}^\mathcal{I}$  of all
smooth maps from the source  supermanifold $\mathcal{I}$ to the
target supermanifold $\mathcal{N}$. (The latter is defined  by
(\ref{N}).) In terms of local coordinates $\phi^k=(\varphi^I,\bar
\varphi_J)$ on $\mathcal{N}$ each element $(\phi:
\mathcal{I}\rightarrow \mathcal{N}) \in \mathcal{N}^\mathcal{I}$
defines (and is defined by) a field configuration
\begin{equation}
\phi^k(z)=\phi^k(t)+\theta \stackrel{\ast}{\phi}{}^k(t)\,.
\end{equation}
According to the definitions above
\begin{equation}\label{}
    \epsilon(\phi^k)=
    \epsilon (\stackrel{\ast}{\phi}{}^k)+1\,,\qquad
    \mathrm{gh}(\phi^k(t))=\mathrm{gh}(\stackrel{\ast}{\phi}{}^k(t))+1=
    \mathrm{gh}(\phi^k)\,.
\end{equation}

The action of the topological sigma-model reads
\begin{equation}\label{tsm}
    \mathcal{S}[\phi]=\int_{\mathcal{I}}d^2z(\Lambda_k(\phi)\mathrm{D}\phi^k-\Omega(\phi))\,.
\end{equation}
The first and second terms are given here by the pull-backs of the
symplectic potential (\ref{Theta}) and the BRST charge
(\ref{bcon}) respectively, and
\begin{equation}\label{}
    \mathrm{D}=\theta\frac{\partial}{\partial t}
\end{equation}
is an odd, nilpotent vector field on $\mathcal{I}$ of ghost number
1. Taking into account the Grassman parity of all the factors
entering the integrand  (\ref{tsm}) one can see that
$\mathrm{gh}(\mathcal{S})=0$. The  action (\ref{tsm}) admits a
straightforward  BV  interpretation that will be given in the next
Sect. \ref{clint}.

The Poisson bracket (\ref{brcov}) on $\mathcal{N}$ induces the
antibracket (i.e. the odd Poisson bracket) on the superspace
$\mathcal{N}^\mathcal{I}$. By definition,
\begin{equation}\label{}
    (F,G)=\int_{\mathcal{I}} d^2z\left(\frac{\delta_r F}{\delta \phi^k(z)}\omega^{km}(\phi(z))
    \frac{\delta_l G}{\delta \phi^m(z)}\right)\,,
\end{equation}
for any functionals of fields $F[\phi]$ and $G[\phi]$.

The model (\ref{tsm}) is called topological since the action
$\mathcal{S}$ is required to  satisfy the classical master
equation
\begin{equation}\label{ss}
    (\mathcal{S},\mathcal{S})=0\,.
\end{equation}
An explicit calculation yields
\begin{equation}\label{}
    \frac 12 (\mathcal{S},\mathcal{S})=\int_{\mathcal{I}} d^2z\left(
    \mathrm{D}\Omega +\frac12\{\Omega,\Omega\}(\phi(z))\right)=\Omega(z_1)-\Omega(z_0)\,.
\end{equation}
To meet the master equation, we impose the following boundary
conditions on the momenta:
\begin{equation}\label{}
    \bar\varphi^J|_{\partial\mathcal{I}}=0\,.
\end{equation}
Then $\Omega(z_0)=\Omega(z_1)=0$, as $\mathrm{Deg}(\Omega)\geq 1$
due to the definition (\ref{16}).

As usual, the classical observables of the topological sigma-model
(\ref{tsm}) are identified with zero-ghost-number cohomology of
the BRST differential $(\mathcal{S},\, \cdot\,)$: The functional
$F$ defines a classical observable iff
\begin{equation}\label{}
\mathrm{gh}(F)=0\quad \mathrm{and}\quad (\mathcal{S},F)=0\,,
\end{equation}
and two BRST closed functionals $F$ and $G$ are considered to
define the same classical observables  ($F\sim G$) if they belong
to the same class of BRST cohomology, i.e. $ F-G=(\mathcal{S}, H)
$ for some $H$.

Of particular interest are the \textit{boundary observables}.
These are constructed from the physical observables $[F]\in
\mathcal{H}_0^0(\mathbb{D})=\mathcal{P}$ of the original gauge
theory (see Sect. \ref{PO}) by the rule $\hat{F}[\phi]=
F(\varphi(z_1))$. A simple computation shows that
\begin{equation}\label{}
    (\mathcal{S},\hat{F})=
    \{\Omega,F\}(\varphi(z_1))=(\mathbb{D}F)(\varphi(z_1))=0\,.
\end{equation}
If  $F=\mathbb{D}G$, then $\hat{F}=(\mathcal{S},\hat{G})$, and
hence $\hat{F}\sim 0$. Clearly, replacing the point $z_1$ with
$z_0$ one gets another set of classical observables supported on
the other end of the superinterval $\mathcal{I}$.

The quantum average of a classical observable $[F]\in
\mathcal{H}_0^0(\mathbb{D})$ corresponding to the original (not
necessarily Lagrangian) gauge theory
$(\mathcal{E},T,d_{\mathcal{E}})$ is defined by the path integral
\begin{equation}\label{average}
    \langle F\rangle = \int_{\mathcal{N}^\mathcal{I}}\mathcal{D}\phi F(\varphi(z_1))e^{\frac i\hbar
    \mathcal{S}[\phi]}\,,
\end{equation}
where the integration measure is normalized in such a way that
$\langle 1\rangle=1$. To evaluate this path integral one has to
impose a gauge fixing condition and choose an appropriate
integration measure. As the sigma-model action entering
(\ref{average}) is a proper solution to the master equation, the
gauge fixing procedure is standard for the BV method \cite{BV},
i.e. this means to fix any Lagrange surface in the anti-Poisson
manifold. The regularization of possible divergencies in
(\ref{average}) does not have any specificity compared to any
other sigma-model path integral.

The main result of the section is Relation (\ref{average})
defining quantum average for a physical observable of any (i.e.
not necessarily Lagrangian) dynamical system in terms of the usual
path integral for the topological sigma-model. Below we elaborate
on the equivalence between the original system and the topological
sigma-model at the level of classical dynamics.

\subsection{Classical equivalence}\label{clint} It was shown in Refs. \cite{GD}, \cite{BM}
that the action of any topological sigma-model on
$(1,1)$-dimensional supermanifold can be interpreted as the BV
master action of a constrained Hamiltonian system. In the case
under consideration such a Hamiltonian system has been already
constructed, in fact, in Sect. 4.6. To recover this effective
Hamiltonian constrained system from the action (\ref{tsm}) one
should integrate out of $\theta$ in (\ref{tsm}) and set to zero
all the fields with nonzero ghost number. The result will have the
form
\begin{equation}\label{S0}
    S_0 [x,\bar x, \lambda]= \int_I (\bar x_i d x^i - \lambda^I\Theta_I(x,\bar
    x))\,.
\end{equation}
where we introduced the notation $\lambda^I\equiv (\stackrel
{\ast}{\bar \eta}{}^a, \stackrel {\ast}{{c}}{}^\alpha )$. Clearly,
$S_0$ is nothing but the Hamiltonian action functional on the
cotangent bundle $T^\ast M$ with the total Hamiltonian given by
the linear combination of the (reducible) first class constraints
(\ref{TRZ}). Upon this identification the 1-forms $\lambda^I$ on
$I$ play the role of Lagrange multipliers to $\Theta$'s. The
action $S_0$ is invariant under the standard gauge transformation
\begin{equation}\label{gtran}
\begin{array}{c}
    \delta_{\varepsilon} x^i=\{x^i,\Theta_I\}\varepsilon^I\,,\qquad \delta_{\varepsilon} \bar
    x_i = \{\bar x_i, \Theta_I\}\varepsilon^I\,,\\[5mm]
    \delta_{\varepsilon} \lambda^I=d\varepsilon^I -\lambda^K
    {U}_{KJ}^I\varepsilon^J +\Xi^I_A\varepsilon^A\,,
\end{array}
\end{equation}
where $\varepsilon^I=(\varepsilon^a,\varepsilon^\alpha)$ and
$\varepsilon^A$ are the infinitesimal gauge parameters, and the
structure functions ${U}_{KJ}^I$, $\Xi^I_A$ are given by
(\ref{ThetaInv}). The compatibility between the gauge
transformations (\ref{gtr}) and the boundary conditions $\bar
x_i(0)=\bar x_i(1)=0$ implies that
$\varepsilon^I(0)=\varepsilon^I(1)=0$. The linear dependence of
constraints $\Theta_I$ leads to the linear dependence of the gauge
algebra generators (\ref{gtran}). Indeed, substituting
\begin{equation}\label{}
\varepsilon^I= \Xi^I_A\varrho^A\,,\qquad  \varepsilon^A=
d\varrho^A - \lambda^I {W}_{IB}^A\varrho^B
\end{equation}
turns (\ref{gtran}) to the trivial (on-shell vanishing) gauge
transformation:
\begin{equation}\label{}
\begin{array}{c}
\displaystyle\delta_{\varrho} x^i=\frac{\delta
S_0}{\delta\lambda^I}\{x^i,\Xi_A^I\}\varrho^A\,,\qquad\delta_{\varrho}
\bar x_i=\frac{\delta S_0}{\delta\lambda^I}\{\bar
x_i,\Xi_A^I\}\varrho^A\,,\\[5mm]\displaystyle
\delta_{\varrho} \lambda^I= -\frac {\delta S_0}{\delta
x^i}\{x^i,\Xi_A^I\}\varrho^A- \frac{\delta S_0}{\delta \bar
x_i}\{\bar x_i,\Xi^I_A\}\varrho^A\,.
\end{array}
\end{equation}
Here we used the definition of the structure function $W$
\begin{equation}\label{}
    \{\Xi^I_B,\Theta_J\}-\Theta_B^K
    {U}_{KJ}^I={W}^{A}_{JB}\Xi^I_A
\end{equation}
following  from the identity $\{\Xi_A^I\Theta_I,\Theta_J\}=0$.

Now the BV interpretation of the sigma-model action functional
$\mathcal{S}=S_0+\cdots$ becomes obvious: it is just the BV master
action corresponding to the theory with reducible first class
constraints and the action (\ref{S0}). Upon this interpretation,
the fields $(\bar\eta^a, \stackrel{\ast}{\bar\xi}{}^A, c^\alpha)$
are identified with the ghosts corresponding to the infinitesimal
gauge parameters $(\varepsilon^a,\varepsilon^A,
\varepsilon^\alpha)$, $\bar\xi^A$ are ghosts for ghosts, and the
other component fields are antifields to the aforementioned ones
including the original gauge fields $(\stackrel {\ast}{\bar
\eta}{}^a, \stackrel {\ast}{{c}}{}^\alpha, x^i,\bar x_i)$.

\  From the Hamiltonian viewpoint, the model (\ref{S0}) has no
(local) degrees of freedom as the first class constraints
$\Theta_I=0$ define a Lagrangian submanifold $L\subset T^\ast M$.
This amounts to saying that, given a ``time''\footnote{To avoid
confusion, let us recall that it is the ``time" $t$  which is an
auxiliary dimension introduced when the original dynamics
(governed by the equations of motion (\ref{T})) is embedded into
the topological sigma-model dynamics. The original equations of
motion (\ref{T}) (that are defined on the boundary, from the
viewpoint of this sigma-model) contain their own evolution
parameter.} moment $t_0\in (0,1)$, by an appropriate gauge
transformation (\ref{gtran}) one can always move any point
$(x^i(t_0), \bar x_j(t_0))\in L$ to any other point of $L$
assigning, simultaneously, any given value to $\lambda(t_0)$, no
matter what were the boundary/initial values of these variables at
$t=0$ or $t=1$. Whereas at the end points of the ``time" interval
we have the boundary conditions $\bar x_i(0)=\bar x_i(1)=0$
reducing the constraints
$$0=\frac{\delta S_0}{\delta \lambda^I}=\Theta_I(x,\bar x)$$
(the only dynamical equations one has actually to solve) to the
original equations of motion $T_a(x(0))=T_a(x(1))=0$. So, we can
conclude that the dynamical content of the topological sigma-model
(\ref{tsm}) is equivalent to that of the original (non-Lagrangian)
gauge theory.

\section{Examples}

\subsection{BV field-antifield formalism}
In this section we quantize the standard Lagrangian gauge system
by the proposed general method that works for not necessarily
Lagrangian theories. In this case, the standard BV quantization
will be shown to follow from the proposed quantization scheme.

In the BV formalism the classical gauge theory is completely
specified by master action $S(\phi)$ defined on an antisymplectic
manifold $\mathcal{M}$ of fields and antifields $\phi^i$ and
subject to the classical master equation
\begin{equation}\label{clasmasteq}
    (S,S)_\mathcal{M}\equiv\frac{\partial_r S}{\partial \phi^I}\sigma^{IJ}\frac{\partial_l S}{\partial
    \phi^J}=0\,,
\end{equation}
$\sigma^{IJ}$ being the odd bivector dual to the antisymplectic
structure $\sigma=d\phi^I\sigma_{IJ}d\phi^J$. Besides the Grassman
parity, the supermanifold $\mathcal{M}$ is graded by the ghost
number and it is additionally required that
\begin{equation}\label{}
    \mathrm{gh}(S)=1\,,\qquad
    \mathrm{gh}((F,G))=\mathrm{gh}(F)+\mathrm{gh}(G)-1\,,
\end{equation}
for any homogeneous $F,G\in C^{\infty}(\mathcal{M})$.

We start with the following simple observation which might be of
some interest in its own right:  \textit{Any Lagrangian gauge
theory on $\mathcal{M}$ with the master action $S(\phi)$ admits an
equivalent reformulation as the topological sigma-model having
$\mathcal{M}$ as target manifold.} The construction is as follows.
Let $\mathcal{J}=\mathcal{I}\times \Pi \mathbb{R}$ be the
$(1,2)$-dimensional supermanifold with boundary, given by the
direct product of the superinterval $\mathcal{I}=\Pi TI$ with
coordinates $z=(t,\theta)$ and the odd linear space $\Pi
\mathbb{R}$ ``parameterized" by $\bar\theta$. The ghost number
assignments are given by
\begin{equation}\label{}
    \mathrm{gh}(t)=0\,,\quad \mathrm{gh}(\theta)=1\,,\quad\mathrm{gh}(\bar\theta)=-1\,.
\end{equation}
The ``points'' of $\mathcal{J}$  are thus the triples
$u=(t,\theta,\bar\theta)$, where $t\in [0,1]$. The boundary of
$\mathcal{J}$ is, by definition, a two point set  $\partial
\mathcal{J}=\{u_0,u_1\}$ constituted by  $u_0=(0, 0, 0)$ and
$u_1=(1,0,0)$.

The field content of the topological sigma-model in question is
identified with the superspace $\mathcal{M}^\mathcal{J}$ of maps
from $\mathcal{J}$ to $\mathcal{M}$. In terms of local coordinates
$\phi^I$ on $\mathcal{M}$, each $\phi \in \mathcal{M}^\mathcal{J}$
is described by a field configuration
\begin{equation}\label{cf}
    \phi^I(u)=\varphi^I(z)+\bar\theta
    \bar\varphi^I(z)=\varphi^I(t)+\theta\stackrel{\ast}{\varphi}{}^I(t)+\bar\theta\bar\varphi^I(t)+\bar\theta\theta
    \stackrel{\ast}{\bar\varphi}{}^I(t)\,,
\end{equation}
 Observe that
under the general coordinate transformations on $\mathcal{M}$,
$\phi^I\rightarrow \tilde{\phi}^I(\phi)$, the component
superfields $\{\bar\varphi^I(z)\}$ behave like the coordinates of
a tangent vector to $\mathcal{M}$. So, one may think of the
superfields $\phi^I(u)=(\varphi^I(z), \bar\varphi^I(z))$ as smooth
maps from the total space of the odd tangent bundle $\Pi TI$ to
that of $\Pi T\mathcal{M}$.

Let us suppose for a moment that the antisymplectic 2-form is
exact i.e. $\sigma=d\rho$ as it happens in the conventional BV
theory. Then we can define the following action of the topological
sigma-model
\begin{equation}\label{S}
\mathcal{S}=\int_{\mathcal{J}}
d^3u(\rho_I(\phi)\bar{\mathrm{D}}\phi^I-S(\phi))\,.
\end{equation}
Here $d^3u=dtd\theta d\bar\theta$ is a natural integration measure
on $\mathcal{J}$,   and
\begin{equation}\label{}
    \bar{\mathrm{D}}=\frac{\partial}{\partial \bar{\theta}}+\mathrm{D}\,,\qquad
    \mathrm{D}=\theta\frac{\partial}{\partial t}
\end{equation}
are odd, self-commuting vector fields on $\mathcal{J}$ of ghost
number 1. Notice that $\mathrm{gh}(\mathcal{S})=0$. As will be
shown below, action (\ref{S}) is equivalent, on the one hand, to
the general topological sigma-model action (\ref{tsm}) constructed
for the original Lagrangian equations of motion with the anchor
being the unit matrix, and on the other hand to the BV master
action for the original Lagrangian theory.

The antibracket on $\mathcal{M}$ gives rise to that on
$\mathcal{M}^{\mathcal{J}}$:
\begin{equation}\label{}
(\phi^I(u),\phi^J(u'))_{\mathcal{M}^\mathcal{J}}=\sigma^{IJ}(\phi(u))\delta^3(u-u')\,.
\end{equation}
Taking the antibracket of the action (\ref{S}) with itself, we
find
\begin{equation}\label{}
    (\mathcal{S},\mathcal{S})_{\mathcal{M}^{\mathcal{J}}}=\int d^3u
    \left( \frac{\delta_r\mathcal{S}}{\delta \phi^I(u)}\sigma^{IJ}(\phi(u))\frac{\delta_l \mathcal{S}}{\delta \phi^J(u)}
    \right)=\frac{\partial \phi^I}{\partial\bar\theta }\left.\left(\sigma_{IJ}\frac{\partial
\phi^J}{\partial\bar\theta}-\partial_IS\right)\right|_{u_0}^{u_1}\,.
\end{equation}
To cancel the boundary terms in the r.h.s. of the last expression
we can set
\begin{equation}\label{}
\left.\frac{\partial\phi^I}{\partial
\bar\theta}\right|_{\partial\mathcal{J}}=0
\quad\Leftrightarrow\quad \bar\varphi^I(0)=\bar\varphi^I(1)=0\,.
\end{equation}
With these boundary conditions the functional $\mathcal{S}$
becomes the classical master action on the antisymplectic
supermanifold $\mathcal{M}^{\mathcal{J}}$. Integrating out of
$\bar\theta$ we get
\begin{equation}\label{SSS}
    \mathcal{S}=\int_{\mathcal{I}} d^2z\left(\bar{\varphi}^I\sigma_{IJ}\mathrm{D}\varphi^I+
    \frac12\bar{\varphi}^I\sigma_{IJ}\bar\varphi^J
    +\bar\varphi^I\frac{\partial S}{\partial\varphi^I}\right)\,.
\end{equation}
As is seen, the functional $\mathcal{S}$ depends actually on the
antisymplectic structure $\sigma$, but not on the choice of
antisymplectic potential $\Lambda$.

In consequence of the master equation the  action (\ref{SSS}) is
invariant under the abelian gauge transformations
\begin{equation}\label{gtr}
    \begin{array}{l}
\displaystyle\delta_\varepsilon \varphi^I=\varepsilon^I\,,\\[3mm]
\displaystyle\delta_\varepsilon \bar
\varphi^J=-\mathrm{D}\varepsilon^J+\varepsilon^L\frac{\partial^2S}{\partial
\varphi^L\partial\varphi^I}\sigma^{IJ}+\varepsilon^L\frac{\partial\sigma_{IK}}{\partial
\varphi^L}(\bar\varphi^K+\mathrm{D}\varphi^K)\sigma^{IJ}\,,
    \end{array}
\end{equation}
with the infinitesimal gauge parameter $\varepsilon^I(z)$ subject
to the boundary conditions
\begin{equation}\label{BC}
    \varepsilon^I(z)\partial_IS(\varphi(z))|_{\partial
    \mathcal{I}}=0\,.
\end{equation}
Since $S$ is a proper solution to the classical master equation
(\ref{clasmasteq}), only ``half'' of the gauge parameters
$\varepsilon^i(z)$ need vanish at the boundary points $z_0$ and
$z_1$.

Notice that the action (\ref{SSS}) is identical in form to the
action (\ref{tsm}). Making identifications
\begin{equation}\label{}
\Lambda =
\bar\varphi^I\sigma_{IJ}d\varphi^J\,,\qquad\Omega=\Omega_1+\Omega_2=
\bar\varphi^I\frac{\partial
S}{\partial\varphi^I}+\frac12\bar{\varphi}^I\sigma_{IJ}\bar\varphi^J\,,
\end{equation}
one can readily check that the BRST charge $\Omega$ obeys the
master equation $\{\Omega,\Omega\}=0$ with the Poisson bracket
corresponding to the exact symplectic structure $\omega
=d\Lambda$. Moreover, when $S$ is the minimal sector master action
of an irreducible gauge theory, the spectrum of component fields
$(\varphi^I, \bar\varphi^J)$ coincides with that presented in
Table 1.  We leave the check of details to the reader.

Notice that the ``truncated'' BRST charge
$\Omega_1=\bar\varphi^I{\partial S}/{\partial\varphi^I}$ was first
considered in Ref.\cite{GD}. In that paper, a very convenient
superfield technique was developed to illuminate the relationship
between the BFV-BRST charge and the BV master action. In this
Section we use a similar technique, although the same results can
be derived without superfields.

According to the results of Sect. 4.2, the term $\Omega_1$ defines
a homological vector field on $\mathcal{M}$:
\begin{equation}\label{}
\begin{array}{c}
\displaystyle QF=\{\Omega_1,F\}=\frac{\partial_r S}{\partial
\varphi^I}\sigma^{IJ}\frac{\partial_l F}{\partial
\varphi^J}\,,\qquad \forall F\in
    C^{\infty}(\mathcal{M})\,,\\[5mm]
    Q^2=0\quad \Leftrightarrow\quad\{\Omega_1,\Omega_1\}=0\,.
    \end{array}
\end{equation}
In the case at hand the operator $Q$ is nothing but the usual
BV-differential associated with the classical master action $S$.
The second term $\Omega_2$ defines (and is defined by) the
original anti-Poisson structure on $\mathcal{M}$:
\begin{equation}
\begin{array}{c}
(F,G)=\{\{\Omega_2,F\},G\}\,, \qquad\forall F,G,H\in
C^{\infty}(\mathcal{M})\,,\\[3mm]
(-1)^{\epsilon(F)\epsilon(H)}(F,(G,H))+cycle(F,G,H)=0\quad\Leftrightarrow\quad
\{\Omega_2,\Omega_2\}=0\,.
\end{array}
\end{equation}
Both structures are compatible in the sense of the graded Liebnitz
rule
\begin{equation}\label{}
    Q(F,G)=(QF,G)+(-1)^{\epsilon (F)+1}(F,QG)\quad
    \Leftrightarrow\quad \{\Omega_1,\Omega_2\}=0\,.
\end{equation}
As we have shown in Sect.\ref{exact}, the last relation implies
the existence of a function $G=G^{IJ}\bar\varphi_I \bar\varphi_J$,
called the propagator, such that
\begin{equation}\label{}
    \Omega_2=\{\Omega_1,G\}\,,\quad \epsilon(G)=0\,,\quad
    \mathrm{gh}(G)=0\,.
\end{equation}
Here we set $\bar\varphi_I\equiv\sigma_{IJ}\bar\varphi^J$, so that
$\epsilon(\bar\varphi_I)=\epsilon(\varphi^I )$ and
$\gh(\bar\varphi_I)=-\gh(\varphi^I )$. When $S(x)$ is the action
of a system without gauge symmetry, the only nonzero block of the
matrix $G^{IJ}$ is given by the inverse to the van Vleck matrix
$\partial_i\partial_j S(x)$ that justifies the term
``propagator''.

To establish a classical correspondence between the topological
sigma-model (\ref{S}) and the original gauge theory we first
observe that according to (\ref{gtr}) the fields $\varphi^I $ are
purely gauge ones in the interior of $\mathcal{I}$, while the
fields $\bar\varphi^I$ enter to the action functional
$\mathcal{S}$ only in an algebraic way (i.e. without derivatives).
This suggests that all the dynamical degrees of freedom are
supported at the boundary of the superinterval $\mathcal{I}$.
Notice also that the form of the action (\ref{SSS}) is quite
similar to that of Hamiltonian action functional with
$\bar\varphi^I$ playing the role of momenta conjugated to the
coordinates $\varphi^I$. So, to obtain the action functional
governing the dynamics of the boundary degrees of freedom we can
just eliminate the auxiliary fields $\bar\varphi^I$ from
$\mathcal{S}$ by means of their own equations of motion:
\begin{equation}\label{99}
    \frac{\delta \mathcal{S}}{\delta
    \bar\varphi^I}=0\quad\Leftrightarrow\quad
    \bar\varphi^I=\mathrm{D}\varphi^I+\sigma^{IJ}\frac{\partial S}{\partial
    \varphi^J}\,.
\end{equation}
As would be expected, the boundary field configurations
$\varphi^I(z_{0,1})$ define the stationary points of $S$, since
$\bar\varphi|_{\partial\mathcal{I}}=\mathrm{D}\varphi|_{\partial\mathcal{I}}=0$.
Substituting (\ref{99}) to (\ref{SSS}), we finally get
\begin{equation}\label{}
\mathcal{S}|_{\delta \mathcal{S}/\delta\bar\varphi=0
}=\int_\mathcal{I}d^2z\mathrm{D}S=S(\varphi(z_1))-S(\varphi(z_0))\,.
\end{equation}
The action describes two uncoupled copies of the original gauge
theory (one for each end of the superinterval $\mathcal{I}$) with
$\mathcal{M}\times \mathcal{M}$ being the total configuration
space.

Let us now comment on quantum equivalence. Proceeding to
quantization, one assigns the BV configuration space $\mathcal{M}$
with a nondegenerate density $\rho$ and replaces the classical
master equation (\ref{clasmasteq}) with the quantum one:
\begin{equation}\label{qmasteq}
    (S,S)_\mathcal{M}=2i\hbar \Delta_{\mathcal{M}} S\,.
\end{equation}
Here $ \Delta_\mathcal{M}:C^{\infty}(\mathcal{M})\rightarrow
C^{\infty}(\mathcal{M})$ is the odd Laplace operator defined by
the rule
\begin{equation}\label{L}
    \Delta_\mathcal{M} F = \mathrm{div}_\rho X_F\,,
\end{equation}
$X_F=(F,\;\cdot\;)_\mathcal{M}$ being the Hamiltonian vector field
corresponding to $F\in C^{\infty}(\mathcal{M})$. The density
$\rho$ is chosen in such a way that $\Delta^2_\mathcal{M}=0$. By
definition, a quantum observable is a function of fields $F(\phi)$
annihilated by the quantum BRST operator $\hat{S}_\hbar$:
\begin{equation}\label{}
    \hat{S}_\hbar F=(S,F)_\mathcal{M}-i\hbar \Delta_\mathcal{M}F=0\,.
\end{equation}
The quantum average of $F$ is defined by the path integral
\begin{equation}\label{}
    \langle F\rangle_S = \int_\mathcal{M}\mathcal{D}\phi \delta(\gamma_a(\phi))F(\phi)e^{\frac
    i\hbar S(\phi)}\,.
\end{equation}
where $\mathcal{D}\phi$ is the integration measure associated to
$\rho$ and equations
\begin{equation}\label{gamma}
\gamma_a(\phi)=0
\end{equation}
define an appropriate Lagrange surface in $\mathcal{M}$. (In the
conventional BV scheme the constraints (\ref{gamma}) are required
to be in abelian involution, i.e.
$(\gamma_a,\gamma_b)_\mathcal{M}=0$, though, upon some
modifications \cite{HyperGauge}, a more general involution is also
allowed.)

As with the antibracket, the measure density $\rho$ on
$\mathcal{M}$ induces that on the space of fields
$\mathcal{M}^{I}$:
\begin{equation}\label{}
    \tilde{\rho}=\prod_{u\in \mathcal{J}}\rho(\phi(u))\,.
\end{equation}
Then the functional counterpart of the odd Laplacian  (\ref{L})
reads
\begin{equation}\label{Lap}
    \Delta_{\mathcal{M}^{\mathcal{I}}}=\int_{\mathcal{J}}d^3u\tilde{\rho}^{-1}
    \frac{\delta}{\delta \phi^I(u)}\tilde{\rho}\sigma^{IJ}(\phi(u)) \frac{\delta}{\delta
    \phi^J(u)}\,.
\end{equation}

Given the quantum master action $S$, we define the action of the
topological sigma-model by the same formula (\ref{S}). The latter
is proved to be  a solution to the quantum master equation on
$\mathcal{M}^{\mathcal{I}}$ with renormalized Plank constant
$\hbar'$. Indeed, a straightforward computation yields
\begin{equation}\label{}
    (\mathcal{S},\mathcal{S})_{\mathcal{M}^{\mathcal{I}}}-2i\hbar'\Delta_{\mathcal{M}^\mathcal{I}}\mathcal{S}=
    \int_\mathcal{M} d^3u \big((S,S)_\mathcal{M}-2i C\hbar'\Delta_{\mathcal{M}}S
    \big)(\phi(u))\,,
\end{equation}
where $C=\delta^3(0)$ is indefinite ``constant". The functional
$\mathcal{S}$ will satisfy the quantum master equation if we set
$\hbar'=\hbar C^{-1}$. Notice that formally $\epsilon(C)=0$ and
$\mathrm{gh}(C)=0$. To assign a precise meaning for the value
$\delta^3(0)$ one has to apply a suitable regularization to the
ill-defined Laplace operator (\ref{Lap}). Similarly, after the
renormalization above any quantum observable $F(\phi)$ of the
original gauge theory gives rise to the boundary quantum
observable $\hat{F}[\phi]=F(\phi(u_1))$ of the topological
sigma-model, i.e. $\hat{\mathcal{S}}_{\hbar'}F(\phi(u_1))=0$.

To calculate the quantum average of a boundary observable
$\hat{F}$ we can apply the Faddeev-Popov recipe to the naive path
integral (\ref{F}). This includes several steps. First one
promotes the infinitesimal gauge parameter $\varepsilon^I(z)$ to a
ghost field $\mathcal{C}^I(z)$ with opposite Grassman parity. Then
one looks for an appropriate gauge fixing conditions. The explicit
structure of the gauge transformations (\ref{gtr}) suggests to
impose  conditions only on the fields $\varphi^I(z)$. This can be
done in many (equivalent) ways. For example, choosing a symmetric
connection $\nabla$ on $\mathcal{M}$ we can set

\begin{equation}\label{}
\chi^I(z)\equiv\frac{\partial \varphi^I}{\partial
\theta}+\theta\big(\ddot\varphi^I-\Gamma^I_{JK}
(\varphi)\dot\varphi^J\dot\varphi^K\big)=0\,.
\end{equation}
Here $\Gamma^I_{JK}$ are the Cristoffel symbols of $\nabla$ and
the overdot stands for the derivative in $t$. Let us assume that
any two points of affine manifold $(\mathcal{M},\nabla)$ are
connected by a unique geodesics. Then equations $ {\chi}{}^I(z)=0$
fix the superfield $\varphi^I(z)$ up to arbitrary boundary values
$\varphi^I(z_0), \varphi^I(z_1)\in \mathcal{M}$. To fix the
residual gauge symmetry at the boundary we may use the conditions
(\ref{gamma}):
\begin{equation}\label{}
\gamma_a(\varphi(z_0))=\gamma_a(\varphi(z_1))=0\,.
\end{equation}
Finally, to provide the correct integration measure in the path
integral (\ref{average}) one introduce the antighost fields
$\bar{\mathcal{C}_J}$. By definition, $\epsilon
(\bar{\mathcal{C}}_I)=\epsilon(\mathcal{C}^I)$. Geometrically, the
fields $\mathcal{C}^I$ are $\bar{\mathcal{C}}_J$ can be viewed as
taking values in tangent and cotangent spaces (with reverse
parity) of the target manifold of fields $\varphi$'s.

Since we regard the ghost-antighosts fields  $\mathcal{C}$'s and
$\bar{\mathcal{C}}$'s to be related with gauge symmetry in the
interior of the superinterval $\mathcal{I}$, the appropriate
boundary conditions for them are
\begin{equation}\label{}
    \mathcal{C}^I|_{\partial \mathcal{I}}=0\,,\qquad
    \bar{\mathcal{C}}_J|_{\partial{\mathcal{I}}}=0\,.
\end{equation}
Notice that we do not introduce ghost and antighost  fields
associated to the residual gauge symmetry at the boundary as such
fields are assumed to be already included into the action $S$ and
the gauges  $\gamma$'s.

After all these preparations we can write
\begin{equation}\label{FPI}
    \langle F\rangle_\mathcal{S}=\int_{\mathcal{M}^{\mathcal{I}}}\mathcal{D}\varphi
\mathcal{D}\bar\varphi
    \mathcal{D}\mathcal{C}\mathcal{D}\bar{\mathcal{C}}\varrho[\varphi]
\delta[\chi(\varphi)]\delta[\gamma(\varphi(z_0))]\delta[\gamma(\varphi(z_1))]F(\varphi(z_1))
    e^{\frac i\hbar \mathcal{S}_{FP}}\,.
\end{equation}
Here
\begin{equation}\label{}
    \mathcal{S}_{FP}[\varphi,\bar\varphi,\mathcal{C},\bar{\mathcal{C}}]=\mathcal{S}[\varphi,\bar\varphi]
    +\mathcal{S}_{\mathrm{gh}}[\varphi,\mathcal{C},\bar{\mathcal{C}}]
\end{equation}
is the usual Faddeev-Popov action given by the sum of the initial
gauge invariant action and the ghost action
\begin{equation}\label{SJ}
\begin{array}{c}
\displaystyle\mathcal{S}_{\mathrm{gh}}\equiv
\int_\mathcal{I}d^2z\,
\bar{\mathcal{C}_I}\big(\delta_{\mathcal{C}}\chi^I\big)\big|_{\chi=0}\\[5mm]
\displaystyle =\int_\mathcal{I} d^2z\left(
\bar{\mathcal{C}}^I\frac{\partial \mathcal{C}_I}{\partial \theta}+
\theta(\nabla_t\mathcal{C}^I\nabla_t\bar{\mathcal{C}}{}_I +
    R_{IJK}^L(\varphi)\mathcal{C}^I\dot\varphi^J\dot\varphi^K\bar{\mathcal{C}}_L)\right)\,,\\[5mm]
    \nabla_t\mathcal{C}^I=\dot{\mathcal{C}}{}^I-\dot\varphi^J\Gamma_{JK}^I(\varphi)\mathcal{C}^K\,,
    \quad     \nabla_t\bar{\mathcal{C}}{}_I=\dot{\bar{\mathcal{C}}}{}_I-\dot\varphi^J\Gamma_{JI}^K(\varphi)
    \bar{\mathcal{C}}{}_K\,.
    \end{array}
\end{equation}
If one disregards the Grassman nature of the Faddeev-Popov ghosts
$\mathcal{C}$'s and $\bar{\mathcal{C}}$'s, then the second term in
(\ref{SJ}) is nothing but the Jacobi action for the deviation of
geodesics. As to the factor
\begin{equation}\label{ro}
  \varrho[\varphi]\equiv
  \mathrm{sdet}\big(\sigma^{IJ}(\varphi(z))\delta^2(z-w)\big)\,,
\end{equation}
it is introduced to provide the invariance of the integration
measure under the gauge transformations (\ref{gtr}). The
appearance of this factor can be rigourously justified within BV
scheme, but we shall not dwell on this here.

Notice that the path integral is Gaussian in the variables
$\bar\varphi$, $\mathcal{C}$ and $\bar{\mathcal{C}}$, and assumes
no actual integration over the interior values of $\varphi$'s due
to the gauge fixing conditions. Integrating successively over all
these variables we get
\begin{equation}\label{F}
\begin{array}{c}
\displaystyle    \langle F
\rangle_\mathcal{S}=\int_{\mathcal{M}\times \mathcal{M}}
\mathcal{D}\varphi(z_0)
\mathcal{D}\varphi(z_1)\delta[\gamma(\varphi(z_0))]
    \delta[\gamma(\varphi(z_1))]F(\varphi(1))e^{\frac{i}{\hbar}(S(\varphi(z_1))-S(\varphi(z_0)))}\\[5mm]
\displaystyle    =(\mathrm{const})\int_{\mathcal{M}}
\mathcal{D}\varphi(z_1)
    \delta[\gamma(\varphi(z_1))]F(\varphi(z_1))e^{\frac{i}{\hbar}S(\varphi(z_1))}\,.
    \end{array}
\end{equation}
(The role of the Faddeev-Popov ghosts was to compensate the
Berezinian resulting from integration of the delta-functional
$\delta[\chi(\varphi)]$ and the factor (\ref{ro}) was exactly
compensated by the integration of $\bar\varphi$'s.) Including the
inessential overall constant in (\ref{F}) to the integration
measure, we arrive at the desired equality
\begin{equation}\label{SM-BV}
    \langle F \rangle_{S}=\langle \hat{F}\rangle_{\mathcal{S}}\,,
\end{equation}
where the $\hat{F}=F(\phi(u_1))$ is the boundary observable of the
topological sigma-model (\ref{S}) constructed from the quantum
observable $F(\phi)$ of the original gauge theory
(\ref{clasmasteq}). The net result, seen from (\ref{SM-BV}), is
that the path integral quantization based on the embedding into
the topological sigma-model (\ref{tsm}) (that does not require the
original equations of motion to be Lagrangian) in the Lagrangian
case brings precisely the same average values for the observables
as in the standard BV-quantization.

\subsection{General Lagrange structure of type (0,0)}
In the previous section we have exemplified the quantization
method applying it to a gauge system whose equations of motion are
Lagrangian. In this section, we apply the method to a
complementary, in a sense, particular case: the system without
gauge symmetry and with independent but general (i.e. not
necessarily Lagrangian) equations of motion. As will be seen, the
quantization method works well in this case too, bringing the
reasonable results admitting clear physical interpretation.

 Let $(\mathcal{E}, T, d_\mathcal{E})$ be a Lagrange structure associated to a set of
independent equations of motion $T_a(x)=0$, so that the matrix
$\partial_i T_a$ is on-shell nondegenerate. According to the
general prescription of Sect.4, this classical theory is BRST
embedded to the Poisson supermanifold. For the sake of simplicity
we assume here that the dynamics bundle $\mathcal{E}$ admits a
flat connection $\nabla=\partial$. The action of the topological
sigma-model has the following structure:
\begin{equation}\label{s4}
\mathcal{S}= \int d^2z \left(\bar x_iDx^i+\bar\eta^a
D\eta_a-\Omega\right)=S_0 + (\mathrm{ghost}\; \mathrm{terms})\,,
\end{equation}
where
\begin{equation}\label{}
S_0=\int_0^1\big(\bar x^i dx^i-\lambda^a \tilde{T}_a (x,\bar
x)\big)\,,\qquad \lambda^a\equiv \overset{*}{\bar\eta}{}^a\,,
\end{equation}
is a Hamiltonian action associated to the first class constraints
(cf. (\ref{TRZ}))
\begin{equation}\label{}
    \qquad \tilde{T}(x,\bar x)=T_a(x)+V_a^i(x)\bar
x_i+O(\bar x^2)\,, \qquad \{\tilde{T}_a,
\tilde{T}_b\}={U}^c_{ab}\tilde{T}_c\,.
\end{equation}
Let us further assume that the classical master action (\ref{s4})
meets also the quantum master equation (\ref{qmasteq}) or, what is
the same, that $\mathcal{S}$ is annihilated by the (suitably
regularized) odd Laplace operator.

In order to write the gauge fixed action we then  introduce the
non-minimal sector of BV fields: the  antighosts $\zeta_a$ and the
Lagrange multipliers $\pi_a$ as well as  the corresponding
antifields  $\overset{*}{\zeta}{}^a$ and
$\overset{\ast}{\pi}{}^a$. The Grassman parity and the ghost
number assignments of the introduced variables  are
\begin{equation}\label{}
\begin{array}{llll}
    \epsilon(\zeta_a)=1\,,\quad&\epsilon(\overset{*}{\zeta}{}^a)=0\,,\quad&
    \epsilon(\pi_a)=0\,,\quad&\epsilon(\overset{*}{\pi}{}^a)=1\,,\\[3mm]
    \mathrm{gh}(\zeta_a)=-1\,,\quad&\mathrm{gh}(\overset{*}{\zeta}{}^a)=0\,,\quad&
    \mathrm{gh}(\pi_a)=0\,,\quad&
    \mathrm{gh}(\overset{\ast}{\pi}{}^a)=-1\,.
\end{array}
\end{equation}
The explicit form of the gauge transformation (\ref{gtran})
suggests to impose the following gauge-fixing condition on
$\lambda$'s:
\begin{equation}\label{gfc}
 \frac{d}{dt}(\lambda^a(e))=0\,,
\end{equation}
with  $e$ being a nowhere vanishing vector field on $[0,1]$. The
gauge fixing fermion associated to (\ref{gfc}) is given by
\begin{equation}\label{gauge fermion}
    \Psi=\int\limits_0^1{dt
    \zeta_a\dot{\lambda}^a}\,,\qquad\gh(\Psi)=-1\,,
\end{equation}
where we set for simplicity $e=\partial_t$. The standard
non-minimal BV action $\mathcal{S}+\int_0^1 dt
\pi_a\overset{*}{\zeta}{}^a$ depends thus  on the fields
$$\phi^A=(x^i,\bar x_i,\lambda^a,\pi_a,\eta_a,
\bar\eta^a,\zeta_a)$$ and antifields
$$\phi^\ast_A=(\overset{*}{x}{}^i,\overset{*}{\bar
x}{}_i,\overset{*}{\lambda}{}^a,\overset{*}{\pi}{}_a,
\overset{*}{\eta}_{a}\overset{*}{\bar\eta}{}^a,\overset{*}{\zeta_a})\,.$$
Now the  gauge fixed action is obtained by restricting the
non-minimal BV action to the Lagrangian submanifold $\mathcal{L}$:
\begin{equation}\label{}
    \phi^{\ast}_A=\frac{\partial \Psi}{\partial \phi^A}\,.
\end{equation}
The last equations allow one to express all the antifields via the
fields. The result is
\begin{equation}\label{gfact}
    \mathcal{S}_{\mathrm{gf}}=\int\limits_0^1{dt\left(\bar x_i\dot{x}^i+
    \pi_a\dot{\lambda}^a+\dot{\bar\eta}{}^a\dot{\zeta}_a+
    \lambda^a\tilde{T}_a(x,\bar x)-\lambda^a\bar\eta^b {U}{}^c_{ab}(x,\bar x)\dot{\zeta}_c\right)}.
\end{equation}

The quantum average of the boundary observable  $F(x(1))$ is
defined now by the regularized version of the naive path integral
(\ref{average})
\begin{equation}\label{Fav}
    \langle F\rangle =\int_{\mathcal{L}} Fe^{\frac i\hbar
    \mathcal{S}_\mathrm{gh}}\,.
\end{equation}

To elucidate the meaning of the last formula it is instructive to
consider the case of trivial Lagrange structure:
\begin{equation}\label{}
    V_{a}^i=0\,,\qquad \tilde{T}_a(x,\bar x)=T_a(x)\,,\qquad {U}_{ab}^c=0\,.
\end{equation}
Integrating in (\ref{Fav}) of $\bar x^i$, $\pi_a$, $\bar\eta^a$
and $\zeta_a$ one finds immediately
\begin{equation}\label{clasamp}
\begin{array}{c}
\displaystyle\langle F \rangle \sim \int
\mathcal{D}x\mathcal{D}\lambda\delta[\dot x^i]\delta[\dot
\lambda^a] F(x(1))e^{\frac i\hbar \int \lambda^a T_a(x)
}\\[5mm]\displaystyle
\sim \int_Md^nx \delta (T_a(x))F(x)\sim F(x_0)
\end{array}
\end{equation}
where $x_0$ is a unique solution to the classical equations of
motion $T_a(x)=0$. Normalizing the integration measure in such a
way that $\langle 1 \rangle=1$, we can finally write   $\langle F
\rangle=F(x_0)$. We see that the quantum average of $F$ involves
no quantum corrections in $\hbar$ and coincides with value of
functional $F(x)$ on a given classical trajectory $x_0\in M$. Thus
one can regards Rel.(\ref{clasamp}) as a classical vacuum-vacuum
amplitude in the presence of observable. In the context of
Hamiltonian mechanics such amplitudes were introduced and studied
in Ref. \cite{Gozzietall}.

In order to relate the above result with conventional formulas of
quantum mechanics let us consider the intermediate possibility:
the anchor is degenerate but regular at the vicinity of a
classical solution $x_0$. In this case, due to Proposition
\ref{LA} we can  assume the equations of motion to have the form
\begin{equation}\label{}
\partial_IS(y)=0\,,\qquad z^J=E^J(y)\,.
\end{equation}
For these equations we have the canonical Lagrange anchor
$V=(V^J,V_I)$, where the vector fields
\begin{equation}\label{}
    V^J=0\,,\qquad V_I=\frac{\partial}{\partial y^I}+\frac{\partial E^J}{\partial
    y^I}\frac{\partial}{\partial z^J}\,
\end{equation}
form an abelian distribution. The integrability conditions
(\ref{TV}) are obviously satisfied with $C$'s equals zero.
Substituting these data to the gauge fixed action (\ref{gfact}),
we arrive at Gaussian path integral for the quantum average
(\ref{Fav}). As in the previous case the path integral is
localized at the boundary. The calculation is  rather simple, so
we just write the final result
\begin{equation}\label{mix}
\langle F \rangle\sim \int dy F(y,E(y)) e^{\frac{i}{\hbar}S(y)}=
\int dydz F(y,z) \delta(z^J-E^J(y))e^{\frac{i}{\hbar}S(y)}\,.
\end{equation}
Here $y^I=y^I(1)$, $z^J=z^J(1)$.  The quantum average is given
thus by a superposition of the classical amplitude (\ref{clasamp})
for the ``non-Lagrangian'' degrees of freedom $z$'s and the usual
Feynman's amplitude associated with the partial action $S(y)$.

In the most general case of irregular Lagrange anchor we can use
the Feynman perturbation technique to obtain the quasi-classical
expansion for the quantum average around the classical solution
$$x_0\in \Sigma\,,\quad \bar x=0\,,\quad  \lambda=0\,,\quad \pi=0\,, \quad \bar
\eta =0\,, \quad\zeta=0\,.$$ Thus we write $x(t)=x_0 +y(t)$, with
a fluctuation field $y(t)$, and decompose the gauge fixed action
on a free part and interaction,
$\mathcal{S}_{\mathrm{gf}}=\mathcal{S}_0+\mathcal{S}_{\mathrm{int}}$,
with
\begin{equation}\label{}
\begin{array}{l}
\displaystyle \mathcal{S}_0=\int\limits_0^1 dt\left(\bar
x_i\dot{y}^i+
    \pi_a\dot{\lambda}^a+\dot{\bar\eta}{}^a\dot{\zeta}_a+\lambda^a\partial_iT_a(x_0)y^i\right)\,,\\[3mm]
\displaystyle \mathcal{S}_{\mathrm{int}}=\int\limits_0^1
dt\left(\lambda^aV_a^i(x_0)\bar
x_i\right.\\[3mm]
\displaystyle
+\lambda^a\sum_{k+l>1}\frac{1}{k!l!}\partial_{i_1}\cdots\partial_{i_k}\partial^{j_1}\cdots\partial^{j_l}\tilde{T}_a(x_0,0)
y^{i_1}\cdots y^{i_k}\bar x_{j_1}\cdots\bar x_{j_l}
\\[5mm]\left.
\displaystyle -\lambda^a\bar\eta^b
\dot{\zeta}_c\sum_{k,l=0}^{\infty}\frac{1}{k!l!}\partial_{i_1}\cdots\partial_{i_k}\partial^{j_1}\cdots\partial^{j_l}
{U}{}^c_{ab}(x_0,0)y^{i_1}\cdots y^{i_k}\bar x_{j_1}\cdots \bar
x_{j_l}\right).
\end{array}
\end{equation}
The Feynman propagator is then deduced from the $\mathcal{S}_0$.
With account of the boundary conditions
\begin{equation}\label{}
\begin{array}{cc}
    \bar x^i(0)=\bar x^i(1)=0\,,\quad &\bar\eta^a(0)=\bar\eta^a(1)=0\,,\\[3mm]
    \pi_a(0)=\pi_a(1)=0\,,\quad&
    \zeta_a(0)=\zeta_a(1)=0\,,
    \end{array}
\end{equation}
we find
\begin{equation}\label{}
\begin{array}{l}
    \langle\lambda^a(t) y^i(s)\rangle_0=i\hbar T^{ai}\,,\\[3mm]\langle \bar
    x_j(t)y^i(s)\rangle_0=i\hbar\delta^i_j[t-\vartheta(t-s)]\,,\\[3mm]
    \langle\bar\eta^b(t)\dot{\zeta}_a(s)\rangle_0=i\hbar\delta_a^b[t-\vartheta(t-s)]\,.
    \end{array}
\end{equation}
Here we use the following definition of $\vartheta$-function
\footnote{There is an unavoidable ambiguity in the definition of
$\vartheta(0)$. The value $\vartheta(0)$ contributes to the path
integral trough the tadpole diagrams involving  $\langle \bar
x_j(t)y^i(t)\rangle_0$ and
$\langle\bar\eta^b(t)\dot{\zeta}_a(t)\rangle_0$. The advantage of
our choice $\vartheta(0)=0$ is that it leads to a covariant
expression for the first quantum correction (\ref{fqcor}). A
similar problem for the path-integral quantization of the Poisson
sigma-model is discussed in Ref. \cite{CaFe}. }:
\begin{equation}\label{}
    \vartheta(t)=\left\{%
\begin{array}{ll}
    1, & \hbox{$t>0$;} \\
    0, & \hbox{$t\leq 0$.} \\
\end{array}%
\right.
\end{equation}

The quasi-classical  expansion for the quantum average of a
boundary observable $F(x(1))$ is given by
\begin{equation}\label{}
    \langle F \rangle = \int F
    e^{\frac{i}{\hbar}\mathcal{S}_{\mathrm{gf}}}=\sum_{n=0}^\infty\frac{i^n}{\hbar^n n!}\int
    F(\mathcal{S}_\mathrm{int})^ne^{\frac{i}{\hbar}\mathcal{S}_{\mathrm{0}}}
\end{equation}
Using the Wick theorem for Gaussian integral we find the following
expression for the first quantum correction to the classical
average:
\begin{equation}\label{fqcor}
    \langle
    F\rangle=F(x_0)+\frac{i\hbar}2\Big[\nabla_i(G^{ij}\partial_jF)-
    \nabla_i(G^{ij}\partial_jT_a)T^{ak}\partial_kF-{U}_{ab}^bT^{ai}\partial_iF\Big](x_0)+O(\hbar^2)\,.
\end{equation}
Here $\nabla$ is some connection on $M$. The symmetric matrix
$G^{ij}\equiv V^i_aT^{aj}$ can be thought of as the Feynman
propagator of boundary fields (cf. (\ref{GGG}))
\begin{equation}\label{}
\langle y^i(t)y^j(s)\rangle= i\hbar G^{ij}(x_0) +O(\hbar^2)
\end{equation}
The expression for the first quantum correction is explicitly
invariant under the general  coordinate  transformations on $M$
and, as one can easily check, does not depend on the choice of
connection $\nabla$.

\subsection{First-order theories}
Let $N$ be a smooth manifold equipped with a vector field
$h=h^{i}(x)\partial_i$. The integral trajectories of $h$ are
defined by the system of first-order ODEs
\begin{equation}\label{v}
T^i(x(t))\equiv \dot{x}^i(t)-h^i(x(t))=0\,,
\end{equation}
where the overdot stands for the derivative in time $t\in
[t_1,t_2]$. To identify these equations with the general Eqs.
(\ref{T})  of Sect.\ref{prelim} one should combine the discrete
index $i$ and the continuous evolution parameter $t$ into the one
superindex $a=(i,t)$. Then the space of histories $M$ is the space
of all smooth trajectories on $N$. Given Eqs. (\ref{v}), we look
for a Lagrange anchor of the form
\begin{equation}\label{VV}
    V^{ij}(t,s)=\alpha^{ij}(x(t))\delta(t-s),
\end{equation}
where $\alpha=\alpha ^{ij}\partial_i\otimes\partial_j$ is a
contravariant tensor on $N$. Substituting the ansatz (\ref{VV})
into the integrability condition (\ref{TV}) we get a set of
necessary and sufficient conditions for the anchor $V$ to be
compatible with equations of motion. These conditions read
\begin{equation}\label{ppp}
 \alpha^{ij}=-\alpha^{ji}\,,\qquad
 [\alpha,\alpha]=0\,,\qquad
        [h,\alpha]=0\,.
\end{equation}
Here the square brackets denote the Schouten  commutator of
multivector fields. Rels. (\ref{ppp}) just say that $\alpha$ is a
Poisson bivector on $N$, and the vector field $h$ is a
differentiation of the corresponding Poisson algebra.

We thus see that the Poisson structure is a particular example of
the Lagrange one.  One recovers it by looking for a local, purely
algebraic anchor (\ref{VV}) for the first-order ODEs (\ref{v}).
When the Poisson bivector $\alpha$ is nondegenerate so is the
anchor $V$. In that case the equations (\ref{v}) appear to be
Hamiltonian and can be derived from a (local) action functional.
Meanwhile, for a degenerate anchor $V$  no such action can exist
even if the equations (\ref{v}) are Hamiltonian.\footnote{ The
last relation in (\ref{ppp}) is automatically satisfied if $h$ is
a (locally) Hamiltonian vector field, i.e.
$h=\rho_i\alpha^{ij}\partial_j$ with $\rho=\rho_i dx^i$ being a
closed 1-form on $N$. For degenerate Poisson bivector the
differentiation $h$, can be not necessarily Hamiltonian even
locally, so the equations (\ref{v}) are more general than the
Hamilton ones.}

In accordance with the definitions of Sect. \ref{rls},  Rels.
(\ref{v}), (\ref{VV}), (\ref{ppp}) define a regular Lagrange
structure of type $(0,0)$. The corresponding BRST charge, being
constructed by the general method of Sect. 4, reads
\begin{equation}\label{BRS}
    \Omega=\int\limits_{t_1}^{t_2}{dt\left(\bar\eta_i(\dot{x}^i-h^i+
    \alpha^{ij}\bar
    x_j)+\frac12\bar\eta_i\bar\eta_j \nabla_k\alpha^{ij}\eta^k\right)},
\end{equation}
Here $\nabla=\partial+\Gamma$ is an arbitrary symmetric connection
$N$.  If we set $\bar\eta_i(t_1)=\bar\eta_i(t_2)=0$, then  the
BRST charge meets the master equation $\{\Omega,\Omega\}=0$ with
respect to the following Poisson bracket:
\begin{equation}\label{brcovv}
\begin{array}{ll}
 \{\bar\eta^i(t),\eta_j(s)\}=\delta_j^i \delta(t-s),&  \{\bar x_i(t), \eta_j(s)\}=\Gamma_{i
 j}^k  \eta_k \delta(t-s) , \\[5mm]
\{\bar x_i(t), x^j(s)\}=\delta_i^j\delta(t-s), &
 \{\bar x_i(t),\bar\eta^j(s)\}=-\Gamma_{ik}^j\bar\eta^k \delta(t-s)\,,\\[5mm]
 \{\bar x_i(t),\bar x_j(s)\}=R_{ijk}^n
\bar\eta_n\eta^k\delta(t-s)\,,&
 \end{array}
  \end{equation}
$R_{ijk}^n(x)$ being the curvature tensor of $\nabla$.
Substituting the BRST charge (\ref{BRS}) and the symplectic
potential for bracket (\ref{brcovv})  into the general formula
(\ref{tsm}), one can get the topological sigma model whose
dynamics is equivalent to the original first order dynamics
(\ref{v}). Having this topological sigma-model, one can compute
the average values (transition amplitudes) for physical
observables by the formula (\ref{average}), even though the
original equations (\ref{v}) are not Hamiltonian.

Notice that the expression (\ref{BRS}) would reproduce the BRST
charge of the Poisson sigma-model if  $h$ was set to zero and
$\nabla=\partial$. This might be viewed as a possible answer to
the question \cite{CaFe} about  the way of incorporating the
Hamiltonian into the Poisson sigma-model. The BRST charge
(\ref{BRS}), containing the covariant derivative of the Poisson
bivector and nilpotent with respect to the non-canonical Poisson
bracket (\ref{brcovv}) probably answers to one more question
\cite{BLN}, about the way of incorporating connection in the
BFV-BRST quantization of Poisson sigma-model to make it explicitly
covariant in the target space. Notice that the ``covariantization"
of BFV scheme was given in \cite{LS2} for general first-class
constraint  systems. The above BRST-BFV formulation for the
topological sigma-model can be viewed as a particular case of the
covariant formalism of the paper \cite{LS2}.

This example allows further extension: the first-order  equations
(\ref{v}) can be complemented by the constraints reducing the
dynamics to a submanifold in $N$. Also the (sub)manifold can be
factorized by a gauge symmetry. In this case, the ansatz
(\ref{VV}) for the Lagrange anchor would result in the requirement
for $\alpha$ to satisfy the Jacobi identity modulo constraints and
gauge generators. Dynamics of this type have been recently studied
in \cite{LS1}, where the deformation quantization method was
extended to such systems. Similar manifolds were also studied in
\cite{CaFe1}. The method of the present paper allows us to define
the transition amplitudes for such systems using the gauged
version of the sigma-model defined by the BRST charge (\ref{BRS}).

\subsection{Maxwell electrodynamics in the first-order
formalism}\label{Max} In this section, we exemplify the general
quantization method by the model of Maxwell electrodynamics in
first-order formalism. This is a simple example, which
demonstrates many characteristic features of more complicated
non-Lagangian field-theoretical models.

In the first-order formalism, the electromagnetic field can be
described by the strength tensor considered as an independent
field, i.e. without use of the electromagnetic potential. The
equations of motion for the strength tensor are not Lagrangian.
These equations are dependent, i.e. there are Noether identities
(\ref{ZT}) among them, but there are no gauge transformations for
the fields. The dynamics bundle (see Remark 4 of Sect. 3.2) is
different from the cotangent bundle, even by dimension. So this
model allows us to exemplify  how the quantization method can
handle with all these features that are impossible in the
Lagrangian dynamics. On the other hand, as the Maxwell
electrodynamics admits alternative Lagrangian formulation
involving the electromagnetic potential, one can check that the
quantization performed by our method gives the same results as in
the standard Lagrangian formalism.

The Maxwell equations for strength tensor $F_{\mu\nu}(x)$ read
\begin{equation}\label{maxeqs}
    T_{1\mu}(x)\equiv\partial^\nu
    F_{\mu\nu}(x)-J_\mu(x)=0\,,\qquad
    T_{2\mu}(x)\equiv\partial^\nu\tilde{F}_{\mu\nu}(x)=0\,,
\end{equation}
where $J_\mu(x)$ is a conserved electric current (considered as an
external source), and
\begin{equation}
\tilde{F}_{\mu\nu}=\frac12\epsilon_{\mu\nu\alpha\beta}F^{\alpha\beta}
\end{equation}
denotes the dual strength tensor. All indices are risen and
lowered by Minkowski metric in $\mathbb{R}^{1,3}$. Due to the
antisymmetry of the strength tensor, Eqs. (\ref{maxeqs}) are
linearly dependent,
\begin{equation}\label{ddT}
    \partial^\mu T_{z\mu}\equiv 0\,,\qquad z=1,2\,.
\end{equation}
To make contact with general notation of the paper introduced in
Section 2, and to identify general relations (\ref{T}), (\ref{ZT})
with (\ref{maxeqs},) (\ref{ddT}) in the Maxwell theory, one should
collect the discrete and continuous indices into the following
superindices: $a=(z,\mu,x)$, $A=(z,x)$ and $i=([\mu\nu],x)$.

Consider the Lagrange anchor $V_a=(V_{1\mu},V_{2\mu})$, where the
Poincar\'e covariant vector fields
\begin{equation}\label{V complete for e-d}
    V_{1\mu}=0\,,\qquad V_{2\mu}=\int{d^4y\partial^\nu\delta^4(x-y)\frac{\delta}{\delta
    \tilde F^{\mu\nu}(y)}}\,.
\end{equation}
form an abelian distribution. Since
\begin{equation}
    V_{2\mu}T_{2\nu}=\frac12(\eta_{\mu\nu}\Box-\partial_\mu\partial_\nu)\delta^4(x-y)\,,\quad
    V_{2\mu}T_{1\nu}=0\,,
\end{equation}
the integrability condition (\ref{TV}) is obviously satisfied. The
anchor is regular but not complete (see Sect. 3.3). The physical
explanation for this incompleteness is that the second equation in
(\ref{maxeqs}), expressing the fact of closedness of 2-form
$F_{\mu\nu}(x)dx^\mu\wedge dx^\nu$, is considered as
non-Lagrangian in the sense of Proposition 2.1. Hence, no quantum
fluctuations violate the condition $dF=0$, and it could be
(locally) resolved in terms of electromagnetic potentials
$A=A_\mu(x)dx^\mu$ both at classical and quantum levels. (Recall
that according to Rel. (\ref{mix}) the non-Lagrangian equations of
motion enter the path integral as arguments of $\delta$-function
thereof suppressing any quantum fluctuations of ``non-Lagrangian''
degrees of freedom).

According to the classification of Sect. 3.2, Rels.
(\ref{maxeqs}), (\ref{ddT}) and (\ref{V complete for e-d})
describe a regular Lagrange structure of type (0,1). The
corresponding BRST charge is given by
\begin{equation}
    \Omega=\int{d^4x\left(\bar\eta^{2\mu}[\partial^\nu
    \tilde F_{\mu\nu}+\partial^\nu\tilde{\bar F}_{\mu\nu}]+\bar\eta^{1\mu}[\partial^\nu F_{\mu\nu}-J_\mu]
    +\bar\xi^z\partial^\mu\eta_{z\mu}\right)},
\end{equation}
$\bar F_{\mu\nu}$ being the momenta canonically conjugated to the
fields $F^{\mu\nu}$. Substituting the BRST charge to  the general
expression (\ref{tsm}) for the sigma-model action yields
\footnote{Both the dynamics bundle and the Noether identity bundle
are assigned with flat connection.}
\begin{equation}
\mathcal{S}=\int\limits_0^1{dtd\theta\left[\Omega+\int{d^4x(\bar
    F_{\mu\nu}DF^{\mu\nu}+\bar\eta^{z\mu}D\eta_{z\mu}
    +\bar\xi^zD\xi_z)}\right]}\,.
\end{equation}
Integrating of $\theta$, we get
\begin{equation}\label{master action reducible}
\begin{array}{c}
\displaystyle
    \mathcal{S}=\int\limits_0^1 dt\int d^4x\Big(\bar F_{\mu\nu}\dot{F}^{\mu\nu}-\bar\eta^{z\mu}\dot{\eta}_{z\mu}+\bar\xi^z\dot{\xi}_z
    +\overset{*}{\bar\eta}{}^{2\mu}[\partial^\nu
    \tilde F_{\mu\nu}+\partial^\nu\tilde{\bar F}_{\mu\nu}]+\overset{*}{\bar\eta}{}^{1\mu}[\partial^\nu
    F_{\mu\nu}-J_\mu]\\[3mm]
    -\bar\eta^{2\mu}\partial^\nu\overset{*}{\tilde{\bar F}}{}_{\mu\nu}-\bar\eta^{2\mu}\partial^\nu\overset{*}{\tilde F}{}_{\mu\nu}-\bar\eta^{1\mu}\partial^\nu\overset{*}{F}{}_{\mu\nu}
    +\bar\xi^z\partial^\mu\overset{*}{\eta}_{z\mu}+\overset{*}{\bar\xi}{}^z\partial^\mu\eta_{z\mu}\Big)\,.
    \end{array}
\end{equation}

The action is invariant under the following gauge transformation:
\begin{equation}\label{rgtr}
\begin{array}{ccc}
\displaystyle    \delta
F^{\mu\nu}=\frac12\varepsilon^{\mu\nu\rho\sigma}\partial_{\rho}\bar\varepsilon^2_{\sigma}\,,\quad&
\displaystyle \delta\bar
    F_{\mu\nu}=\frac12\partial_{[\mu}\bar\varepsilon^1_{\nu]}+\frac12\varepsilon_{\mu\nu\rho\sigma}\partial^\rho\bar\varepsilon^{2\sigma}\,,
    \quad&\delta\overset{*}{\bar\eta}{}^{z\mu}=-\dot{\bar\varepsilon}^{z\mu}+\partial^\mu\overset{*}{\bar\varepsilon}{}^z\,,\\[3mm]
    \displaystyle
    \delta\bar\eta^{z\mu}=-\partial^\mu\bar\varepsilon^z\,,\quad&\delta\overset{*}{\bar\xi}{}^z=-\dot{\bar\varepsilon}^z\,,\quad&\delta\bar\xi^z=0\,.
    \end{array}
\end{equation}
(Here we write out only the transformation formulas for fields.)

\  From the general viewpoint,  (\ref{master action reducible}) is
a minimal master action of a gauge theory with linearly dependent
gauge-algebra generators. The fields $\bar\eta$'s and
$\overset{\ast}{\bar \xi}$'s are ghosts associated to the
reducible gauge symmetry, and the fields $\bar \eta$'s play the
role of ghosts-for-ghosts.

 The gauge fixing procedure for the
theory at hands is standard \cite{HT}. We introduce the
``non-minimal sector of variables'' constituted by the trivial
field-antifield pairs  ($b$'s, $\overset{\ast}{b}$'s) and
($\pi$'s, $\overset{\ast}{\pi}$ 's). The spectrum, parity and
ghost number of these fields are completely determined by the form
of gauge transformations (\ref{rgtr}) (see \cite{HT} for details).
An appropriate gauge-fixing fermion is chosen as
\begin{equation}
    \Psi=\int\limits_0^1{dt\int d^4x\big(b_{z\mu}\dot{\lambda}^{z\mu}+\dot{b}_z\partial_\mu\lambda^{z\mu}+b_{1z}
    \partial_\mu\bar\eta^{z\mu}+b^{1z}_1\partial^\mu\dot{b}_{z\mu}\big)}\,,\qquad\gh(\Psi)=-1\,,
\end{equation}
where we set $\lambda^{z\mu}\equiv
\overset{*}{\bar\eta}{}^{z\mu}$. The gauge-fixed action is
obtained by excluding the antifields from  the non-minimal master
action
\begin{equation}
    S+\int\limits_0^1{dt\int d^4x\big(\pi_{z\mu}\overset{*}{b}{}^{z\mu}+\pi_z\overset{*}{b}{}^z
    +\pi_{1z}\overset{*}{b}{}^z_1+
    \pi^{1z}_1\overset{*}{b}{}^1_{1z}\big)}\,.
\end{equation}
using the equation  $\phi^\ast_i={\delta\Psi}/{\delta \phi^i}$,
where $\phi^i$ and $\phi^*_i$ collectively denote all the fields
and antifields respectively. The result is
\begin{equation}\label{action gauge fixed for ed}
\begin{array}{c}
\displaystyle     \mathcal{S}_{\mathrm{gf}}=\int
d^4x\Big\{\int\limits_0^1dt\Big[\bar
F_{\mu\nu}\dot{F}^{\mu\nu}+\bar\eta^{z\mu}(\ddot{b}_{z\mu}+\partial_\mu\ddot{b}_z)+\lambda^{2\mu}(\partial^\nu
    \tilde F_{\mu\nu}+\partial^\nu\tilde{\bar F}_{\mu\nu})+\lambda^{1\mu}(\partial^\nu
    F_{\mu\nu}-J_\mu)\\[3mm]\displaystyle+b_{1z}\Box\bar\xi^z
    -(\partial^\mu\dot{b}_{z\mu}+\Box\dot{b}_z)\overset{*}{\bar\xi}{}^z
    +\pi_{z\mu}(\dot{\lambda}^{z\mu}+\partial^\mu\dot{b}^{1z}_1)-\pi_z\partial_\mu\dot{\lambda}^{z\mu}+\pi_{1z}\partial_\mu\bar\eta^{z\mu}+\pi^{1z}_1\partial^\mu\dot{b}_{z\mu}\Big]
    \\[3mm]
\left.
+\left(\dot{\bar\eta}^{z\mu}b_{z\mu}+b_{z\mu}\partial^\mu\overset{*}{\bar\xi}{}^z+\pi_{z\mu}\partial^\mu
b^{1z}_1+\pi_z\partial_\mu\lambda^{z\mu}\right)\right|_{0}^{1}\Big\}\,.
    \end{array}
\end{equation}
We impose the following boundary conditions at $t=0,1$:
\begin{equation}
    \bar\eta^{z\mu}=0\,,\quad\bar\xi^z=0\,,\quad\pi^{1z}_1=0\,,\quad b_{z\mu}=-\partial_\mu
    b_z\,,\quad b^{1z}_1=0\,.
\end{equation}

The physical observables are just arbitrary functionals of the
strength tensor $F_{\mu\nu}(x)$ and the quantum average $\langle
\mathcal{O}\rangle$ of an observable $\mathcal{O}(F_{\mu\nu})$ is
given by the general expression (\ref{Fav}). As with an arbitrary
free theory, the ghost dynamics is completely decoupled from the
dynamics of physical fields. Integrating over the interior values
of ghost fields we arrive at the path integral with action
$$
    \mathcal{S}_{\mathrm{gf}}=\int d^4x\Bigl\{\int\limits_0^1dt\left[\bar{F}_{\mu\nu}\dot{F}^{\mu\nu}+A^{2\mu}(\partial^\nu
    \tilde F_{\mu\nu}+\partial^\nu\tilde{\bar F}_{\mu\nu})+A^{1\mu}(\partial^\nu F_{\mu\nu}-J_\mu)\right]
    +\left.\left(b_z\Box\overset{*}{\bar\xi}{}^z+\pi_z\partial_\mu
    A^{z\mu}\right)\right|_{0}^1\Bigr\},
$$
where $A^{z\mu}(x)\equiv\lambda^{z\mu}(x,0)$. (One can check that
all the determinants resulting from the integration are cancelled
out.) Then, integrating of the remaining variables $F_{\mu\nu}$,
$\bar{F}_{\mu\nu}$, $A^{z\mu}$, $b_z$, $\overset{*}{\bar\xi}{}^1$
and $\pi_z$, we arrive at the final result
\begin{equation}
\langle\mathcal{O}\rangle=(\mathrm{const})\int
\mathcal{D}A\mathcal{D}F\mathcal{D}\pi
\mathcal{D}c\mathcal{D}\bar{c}\,\mathcal{O}(F_{\mu\nu})e^{\frac{i}{\hbar}
\tilde{S}_{\mathrm{gf}}}\,. \label{maxaverage}
\end{equation}
Here we introduced the following notation:
\begin{equation}\label{fin}
    \tilde{S}_\mathrm{gf}=\int d^4x \Big(F^{\mu\nu}[\partial_\mu A_{\nu}+F_{\mu\nu}]+\pi\partial^\mu
    A_{\mu}+\partial_\mu c\partial^\mu\bar c \Big)\,,
\end{equation}
\begin{equation}\label{}
    A_\mu =A_{1\mu}(x)\,,\quad F_{\mu\nu}=F_{\mu\nu}(x,1)\,,\quad \pi=\pi_1(x,1)\,,\quad
    c=\overset{\ast}{\bar\xi}{}^1(x,1)\,,\quad \bar c= b_1(x,1)\,.
\end{equation}
As is seen the functional (\ref{fin}) is nothing but the
gauge-fixed action of Maxwell's electrodynamics in the first-order
formalism: The boundary values of the Lagrange multipliers
$\lambda_{1\mu}$ are identified with electromagnetic potentials
$A_{\mu}$, whereas the boundary values of ghost-for-ghost
$\overset{\ast}{\bar\xi}{}^1$ and $b_1$ play the role of the
Faddeev-Popov ghosts associated with the gauge invariance
$A_\mu\rightarrow A_\mu+\partial_\mu\varphi$ and the Lorentz gauge
$\partial^\mu A_\mu$.

Above, one can see how the non-Lagrangian Maxwell's equations for
the strength tensor (\ref{maxeqs}) are quantized strictly
following the general prescription of the paper, through
quantizing equivalent topological field theory in five dimensions
(\ref{master action reducible}). After integrating of the fields
in the bulk (that can be done explicitly in this case) the result
reduces to the standard expression (\ref{maxaverage}) for the
quantum average given by the Faddeev-Popov quantization of
electrodynamics in terms of electromagnetic potential.

\section{Concluding remarks} In this paper we have proposed a
method of converting any not necessarily Lagrangian dynamics into
equivalent Lagrangian topological field theory, that allows us to
path-integral quantize the dynamical system. The Lagrange
structure described in Sections 2 and 3 is a key pre-requisite of
this quantization method. Given the equations of motion (\ref{T}),
they completely define their gauge symmetries (\ref{RT})  and the
Noether identities (\ref{ZT}), that are the basic ingredients for
constructing the Lagrange structure. The Lagrange anchor
(\ref{TV}) is the last input the Lagrange structure needs to be
defined. This ingredient can not be uniformly found from the
equations of motion, as the compatibility conditions (\ref{TV})
are not so restrictive to determine the anchor in a unique way.
This is much similar to the Poisson structure which is not
uniquely defined by given first order equations of motion (As is
seen from the example of Sect. 6.3, this is more than a general
analogy). Different Lagrange anchors can result in different
quantization for the same classical theory. The trivial anchor
$V=0$, which is compatible with any equations of motion, results
in a trivial quantization in the sense that the quantum average of
any physical observable would coincide with its classical value.
Choosing a degenerate Lagrange anchor of constant rank, one
implicitly separates the degrees of freedom which are quantized
from those which are not. This is a choice which can have physical
and/or geometrical motivation, but it can not be justified just by
the form of the equations of motion. Even in the simplest case
with equations of motion following from the action with no gauge
symmetry, one can choose, in principle, not a unit anchor (that
would result in the standard Feynman path integral) but a
degenerate one that would assign no quantum corrections to some
degrees of freedom. And this latter choice might have reasonable
physical motivation in some cases, e.g. when some ``part" of
dynamics is to be considered as an effective theory emulating
classical background for the other part that demonstrates a
quantum behavior.

So, the general conclusion is that the Lagrange structure which is
behind the path integral quantization, requires more
physical/geometrical data about the system than it is contained in
the classical equations of motion. This is not surprising: as
quantum theory gives a more detailed description of the model, it
has to require more inputs.

There are many physically interesting models, like self-dual
Yang-Mills theory, Vasiliev's higher-spin fields \cite{Va}, etc.,
having no Lagrangians. These models can have, however, nontrivial
Lagrange structures that could allow to quantize these
non-Lagrangian theories. As the equations of motion are not
sufficient, in general, to uniquely define the Lagrange anchor,
one has to identify in these models some other appropriate
structures that could serve as building blocks for constructing
the anchor. There are various ideas that can be helpful for making
such an identification in local field theories. For example, not
so many explicitly relativistic covariant finite order
differential operators can be found for a given field, so even the
most general ansatz for a local explicitly covariant Lagrange
anchor can be amenable to study in many cases. As is seen from the
example of section 6.4, there are quite few first order operators
for the abelian vector field that can be tested for this role. If
the space of trajectories $M$ admits any symmetric contravariant
second rank tensor $G^{ij}$ (degenerate or not, no matter) even
not related anyhow with the equations of motion, it can be taken
as a propagator of Sect. 4.4, that would define the Lagrange
anchor in the form $V_a^i= \partial_j T_a G^{ji}$. Even the
requirement of symmetry can be relaxed for $G^{ij}$ to hold on
shell only. If other geometric data are naturally assigned to the
system, other schemes can be invented, perhaps, to convert these
data into the Lagrange anchor. In this paper we just propose the
method to path integral quantize dynamics, given equations of
motion and Lagrange anchor.

\end{document}